\documentclass[reprint,prb,aps,amsmath,amssymb,floatfix,superscriptaddress,longbibliography]{revtex4-1}
\usepackage{dcolumn}
\usepackage{bm}
\usepackage{graphicx}
\usepackage{amsthm}
\usepackage{color}
\usepackage{xcolor}
\usepackage{verbatim}
\usepackage{esint}
\usepackage[english]{babel}
\usepackage{amsfonts}
\usepackage{slashed}
\usepackage{latexsym}
\usepackage{bm}
\usepackage[colorlinks = true,
            linkcolor = red,
            urlcolor  = blue,
            citecolor = magenta,
            anchorcolor = red]{hyperref}
\usepackage{esint}
\usepackage{soul}
\usepackage{cancel}
\usepackage[normalem]{ulem}
\usepackage{empheq}
\usepackage{dsfont}
\usepackage{amsmath}
\usepackage{ulem}
\usepackage{epsfig}
\usepackage{enumerate}
\definecolor{sapphire}{rgb}{0.03, 0.03, 0.41}

\DeclareMathAlphabet\mathbfcal{OMS}{cmsy}{b}{n}
\def\be{\begin{equation}}
\def\ee{\end{equation}}

\begin{document}

\title{Minimal two-body quantum absorption refrigerator}

\author{Bibek Bhandari}
\affiliation{Department of Physics and Astronomy, University of Rochester, Rochester, NY 14627, USA}
\author{Andrew N. Jordan}
\affiliation{Department of Physics and Astronomy, University of Rochester, Rochester, NY 14627, USA}
\affiliation{Center for Coherence and Quantum Optics, University of Rochester, Rochester, NY 14627, USA}
\affiliation{Institute for Quantum Studies, Chapman University, Orange, CA 92866, USA}




             
\begin{abstract}
We study the phenomenon of absorption refrigeration, where refrigeration is achieved by heating instead of work, in two different setups: a minimal set up based on coupled qubits, and two non-linearly coupled resonators. Considering ZZ interaction between the two qubits, we outline the basic ingredients required to achieve cooling. Using local as well as global master equations, we observe that inclusion of XX type term in the qubit-qubit coupling is detrimental to cooling. We compare the cooling effect obtained in the qubit case with that of non-linearly coupled resonators (multi-level system) where the ZZ interaction translates to a Kerr-type non-linearity. For small to intermediate strengths of non-linearity, we observe that multi-level quantum systems, for example qutrits, give better cooling effect compared to the qubits. Using Keldysh non-equilibrium Green's function formalism, we go beyond first order sequential tunneling processes and study the effect of higher order processes on refrigeration. We find reduced cooling effect compared to the master equation calculations.
\end{abstract}

\maketitle

\section{Introduction}
\label{sec:intro}
Minimal heat engines and refrigerators based on quantum systems offer the prospect for future thermal devices. Understanding thermal transport in quantum systems along with subsequent realization of corresponding thermal devices has been the focus of intense research for some time \cite{scovil1959,geusic1967,pendry1983,geva1992,
allahverdyan2000,kieu2004,
hanggi2009,horodecki2013,benenti2017,
anders2016,alicki1979,yang2020}. This effort has been further fueled by experimental advancement\cite{schwab2000,meschke2006,blickle2012,
brantut2013,koski2013,martinez2016,
rossnagel2016,thierschmann2015,josefsson2018} in the field giving new insights about heat flow in quantum thermal devices.

\begin{figure}[!htb]
\includegraphics[width=0.9\columnwidth]{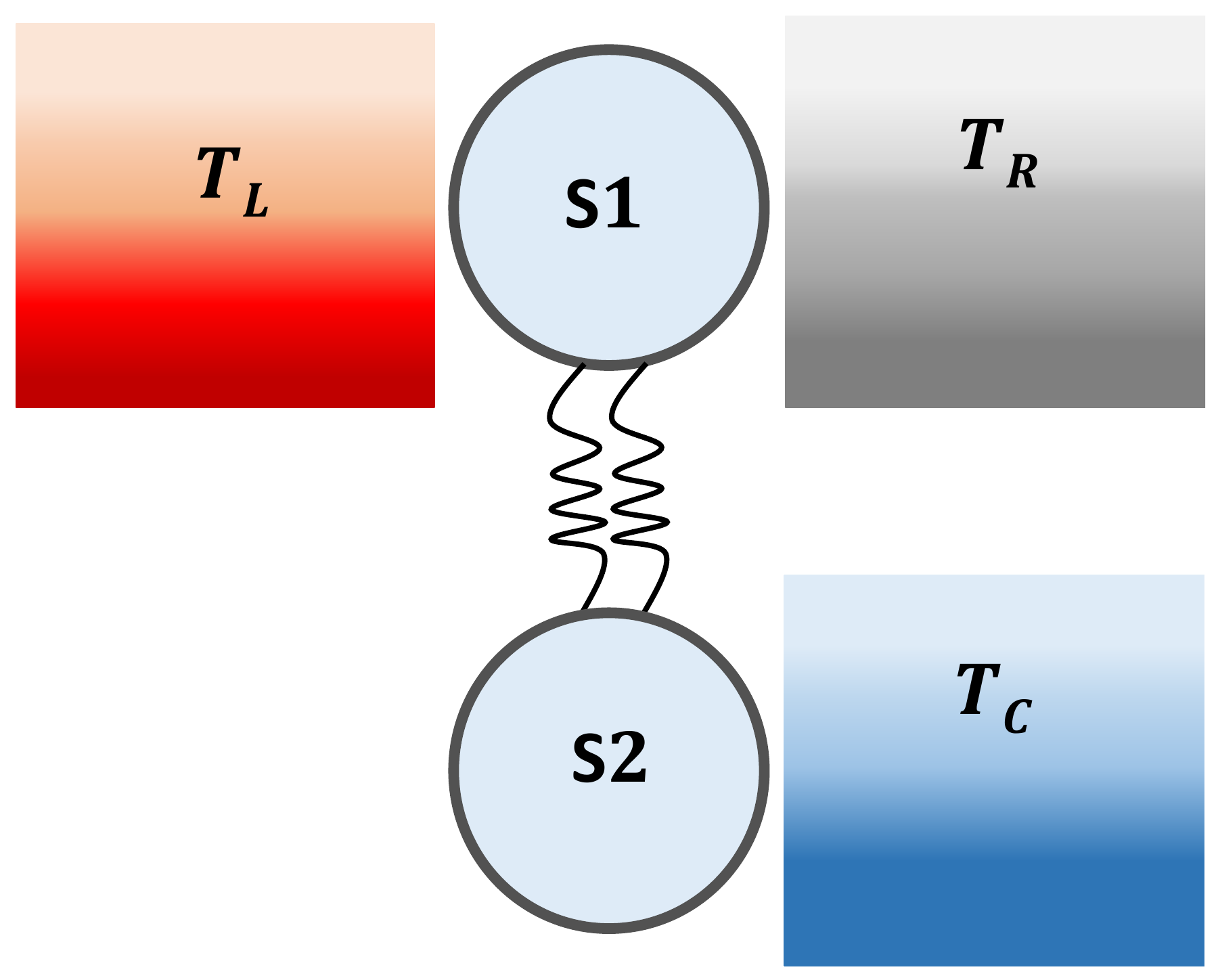}	
	\caption{Schematic representation of the setup. The two systems S1 and S2 can either be qubits or resonators. The system S1 is connected to two baths L and R which are kept at temperatures $T_{\rm L}$ and $T_{\rm R}$ whereas the system S2 is coupled to the bath C which is kept at temperature $T_{\rm C}$, such that $T_L>T_R\geq T_C$. The baths have Lorentzian spectral density as defined in Eq.~(\ref{eq:spectral_den_lor}). }
	\label{fig:setup}
\end{figure}

In this article, we will focus on absorption refrigeration in minimal quantum setups. Absorption refrigeration refers to the phenomenon of cooling a cold bath by maintaining a thermal bias across other two baths, see Fig.~\ref{fig:setup}. Intriguingly, the refrigeration process is driven by heating the hot bath. The phenomenon has been studied theoretically in a variety of setups ranging from quantum dots\cite{benenti2017,paolo2018,
cleuren2012}, electronic cavities\cite{manikanandan2020,
etienne2021}, superconducting systems\cite{marchegiani2018,
pekola2007}, trapped ions\cite{maslennikov2019} to qubits\cite{linden2010,
skrzypczyk2011,brunner2012,
brask2015,mitchinson2015,
silva2015,he2017,hewgill2020,
naseem2020,hammam2021,
manzano2019} and resonators\cite{naseem2020,kosloff2011}. However, to our knowledge, the only experimental work on absorption refrigeration was done in Ref.~\onlinecite{maslennikov2019} for the case of trapped ions. 

The absorption refrigerators based on two-body interactions are studied both for electronic\cite{paolo2018} and bosonic systems\cite{naseem2020,linden2010}. The smallest possible two-body refrigerator proposed by Linde et al in Ref.~\onlinecite{linden2010} consisted of a coupled qubit-qutrit system. Recently, it was noticed that refrigeration can also be achieved with anisotropically coupled qubits\cite{naseem2020}. In this article, we demonstrate that the absorption refrigeration can be obtained in a simpler setup, namely a qubit-qubit system with ZZ coupling. Notably, ZZ-coupling between two qubits has been experimentally realized in circuits ranging from qubits based on quantum dots to Josephson junctions\cite{dots,super}. An electronic version of our set up based on Coulomb coupled quantum dots was studied in Ref.~\onlinecite{paolo2018}. However, a natural question arises regarding if the setup based on qubits is the best refrigerator. We will address this question by investigating refrigeration in non-linearly coupled resonators (multi-level system).

To obtain absorption refrigeration in small setups, selective transport is the key. There are different ways to introduce selective transport - one of the main approach which we will consider in this article is to introduce energy filters in the contact region\cite{paolo2018,
manikanandan2020,sanchez2011}. Most of the previous works on quantum absorption refrigeration are based on single photon transition (sequential) processes with a sharp energy requirement. In this regime, suitable engineering of the device can lead to highly selective transport. But higher order processes which include two or more photons can decrease the effectiveness of selective transport by broadening the energy window for transport. These processes become more important when the system-bath coupling is not weak enough. Seemingly, the effect of higher order processes on absorption refrigeration calls for research.

The paper is organized as follows. In the next section, we will propose a model for obtaining absorption refrigeration based on coupled qubits or resonators. In Sec.~\ref{sec:qubit}, we will present a detailed study of absorption refrigerators based on two qubits with ZZ coupling. In Sec.~\ref{sec:zzxx}, we further our investigation by considering a more generic qubit-qubit coupling condition. Particularly, we take XX as well as ZZ coupling between the qubits and use both  local as well as global master equation to study the dynamics. In Sec.~\ref{sec:res}, we will study absorption refrigeration in two non-linearly coupled resonators. We will also compare the cooling effects in the aforementioned two setups. Finally in Sec.~\ref{sec:resgreen}, we will use the Keldysh nonequilibrium Green's function formalism to study absorption refrigeration in the case of coupled resonators going beyond first order sequential tunneling processes.

\section{Model}
The setup under investigation, depicted in Fig.~\ref{fig:setup} consists of two coupled systems, namely $S_1$ and $S_2$. The system $S_1$ is attached to two baths: one of them is kept at temperature $T_H=T+\Delta T$ and the other one at temperature $T$. The system $S_2$ has only one bath attached to it which is kept at a colder temperature $T_{\rm C}$.
The total Hamiltonian for the set up is given by
\begin{equation}
H=H_{S}+H_B+H_C,
\end{equation}
where $H_S$ represents the Hamiltonian for the system. We shall consider two different types of systems: 1) coupled qubits (Q) and 2) non-linearly coupled resonators (R). Further, we consider bosonic baths (B) with Hamiltonian
\begin{equation}
H_B=\sum_{k,\alpha}\epsilon_{k\alpha}b_{k\alpha}^\dagger b_{k\alpha},
\label{eq:ham_bath}
\end{equation}
where $\epsilon_{k\alpha}$ is the energy of mode $k$ of bath $\alpha$. $b_{k\alpha}(b_{k\alpha}^\dagger)$ are the annihilation (creation) operators for bath $\alpha$. The contact Hamiltonian $(H_C)$ depends on whether the system is taken to be a qubit or a resonator. 
It has been observed in several previous works that energy filtering plays a significant role in achieving refrigeration\cite{paolo2018,
manikanandan2020,sanchez2011}. For that purpose, we take a harmonic resonator between the baths and the system which helps in energy filtering. The coupling strength between the bath $\eta$ and the resonator is taken to be $\gamma_\eta$ and the one between the resonator and the system as $\Gamma_\eta$. The frequency of the resonator attached to the bath $\eta$ is $\Omega_\eta/\hbar$. We consider $\Gamma_\eta \ll \gamma_\eta$, such that the bath-resonator acts as an effective bath with a Lorentzian-type spectral density
\begin{equation}
K_\eta(\omega)=\frac{\Gamma_\eta(\omega)}{(\omega-\Omega_\eta)^2+\Gamma_\eta^2(\omega)},
\label{eq:spectral_den_lor}
\end{equation}
where $\eta={\rm L,R,C}$ and $\Gamma_\eta(\omega)=\Gamma_\eta \omega \Theta(\omega-\epsilon_{\rm c})$, $\epsilon_{\rm c}$ being the cut-off frequency. The energy filters allow photons with energy equal or close to the resonant energy of the resonator to pass from the system to the baths and vice-versa. The energy window available for transport is determined by the width of the Lorentzian given by the system-bath coupling strength$(\Gamma_\eta(\omega))$. Hence, the degree of filtering depends on the system-bath coupling strength as well as the details of the system. When the system is discrete and the adjacent levels are well separated, the filtering effect is strong provided system-bath coupling is weak enough. The filters become less efficient when more than one level exists within the energy window provided by the system-bath coupling strength and vanishes for systems with continuous degrees of freedom.
\section{Minimal absorption refrigerator with two qubits}
\label{sec:qubit}
In this section, we will present a minimal model for the absorption refrigerator based on a coupled qubit system. We consider the set up of Fig.~\ref{fig:setup}, where we will take a qubit (Q1) as system S1 and another qubit (Q2) as system S2. The Hamiltonian for the coupled qubit system is given by
\begin{equation}
H_S=\sum_{n=0,1,2}\epsilon_n{\pi}_{nn}
+\epsilon_d{\pi}_{dd}+\Delta_x\left({\pi}_{21}+{\pi}_{12}\right), 
\label{eq:sysqbt}
\end{equation}
where the projection operator on the system states ${\pi}_{mm}=|m\rangle\langle m|$, $m$ represents different possible sates of the coupled qubit system: $0$ when both the qubits are in the ground state, $1$ when Q1 is in the excited state, $2$ when Q2 is in the excited state and $d$ when both the qubits are in the excited state. $\epsilon_n$ is the energy associated with the corresponding state $n$. For the sake of simplicity, we consider $\epsilon_0=0$. In the presence of ZZ interaction between the two qubits, $\epsilon_d=\epsilon_1+\epsilon_2+\Delta_z$ where $\Delta_z$ gives the strength of ZZ interaction. Finally, the last term on the right hand side gives the $XX$ interaction between the two qubits. The contact Hamiltonian is given by
\begin{multline}
H_{ C}=\sum_{k,\alpha} V_{k\alpha}\left({\pi}_{01}+{\pi}_{d2}+h.c.\right)(b_{k\alpha}+b_{k\alpha}^\dagger)\\
+V_{kC}\left({\pi}_{02}+{\pi}_{d1}+h.c.\right)(b_{kC}+b_{kC}^\dagger),
\end{multline}
where $\alpha={\rm L,R}$.
The baths are bosonic baths as given in Eq.~(\ref{eq:ham_bath}). Moreover,  we shall consider Lorentzian spectral density for the baths as given in Eq.~(\ref{eq:spectral_den_lor}).  We shall use both the local and global {\em master equations} to study the energy dynamics (see App.~\ref{app:mas} for details). Although for the sake of completeness, we keep both rotating and counter rotating terms in the contact Hamiltonian, the master equation takes into account only the contributions coming from the rotating terms.
\subsection{ZZ coupling}
\label{sec:zz}
In this section, we will discuss the mechanism as well as the necessary conditions to cool the coldest bath in the absence of XX coupling ($\Delta_x=0$). We can describe the state of the coupled qubit system by assigning probabilities to different states. Hence, we assign $p_0,p_1,p_2,p_d$ for the states $0,1,2,d$ respectively. In order to calculate the master equation, we need to define different possible transition rates. We consider weak system-bath coupling such that only one photon processes are allowed.
Let $\Gamma_{lm}=\sum_{\alpha}\Gamma_{lm,\alpha}$ be the total transition rate from state $l$ to state $m$ corresponding to all possible baths $\alpha$. The transition rates can be broadly categorized into outgoing rates (transitions which take one of the qubit to ground state) and incoming rates (transitions which take one of the qubit to excited state). The outgoing and incoming transition rates corresponding to same qubit states and same bath satisfy detailed balance equation. For instance,
\begin{equation}
\Gamma_{lm,\alpha}^{(o)}=  e^{(\epsilon_l-\epsilon_m)/k_{\rm B}T_\alpha}\Gamma_{ml,\alpha}^{(i)},
\end{equation}
gives the relation between the outgoing (o) rate $\Gamma_{lm,\alpha}$ and the incoming (i) rate $\Gamma_{ml,\alpha}$. 
\begin{figure}[!htb]
\includegraphics[width=\columnwidth]{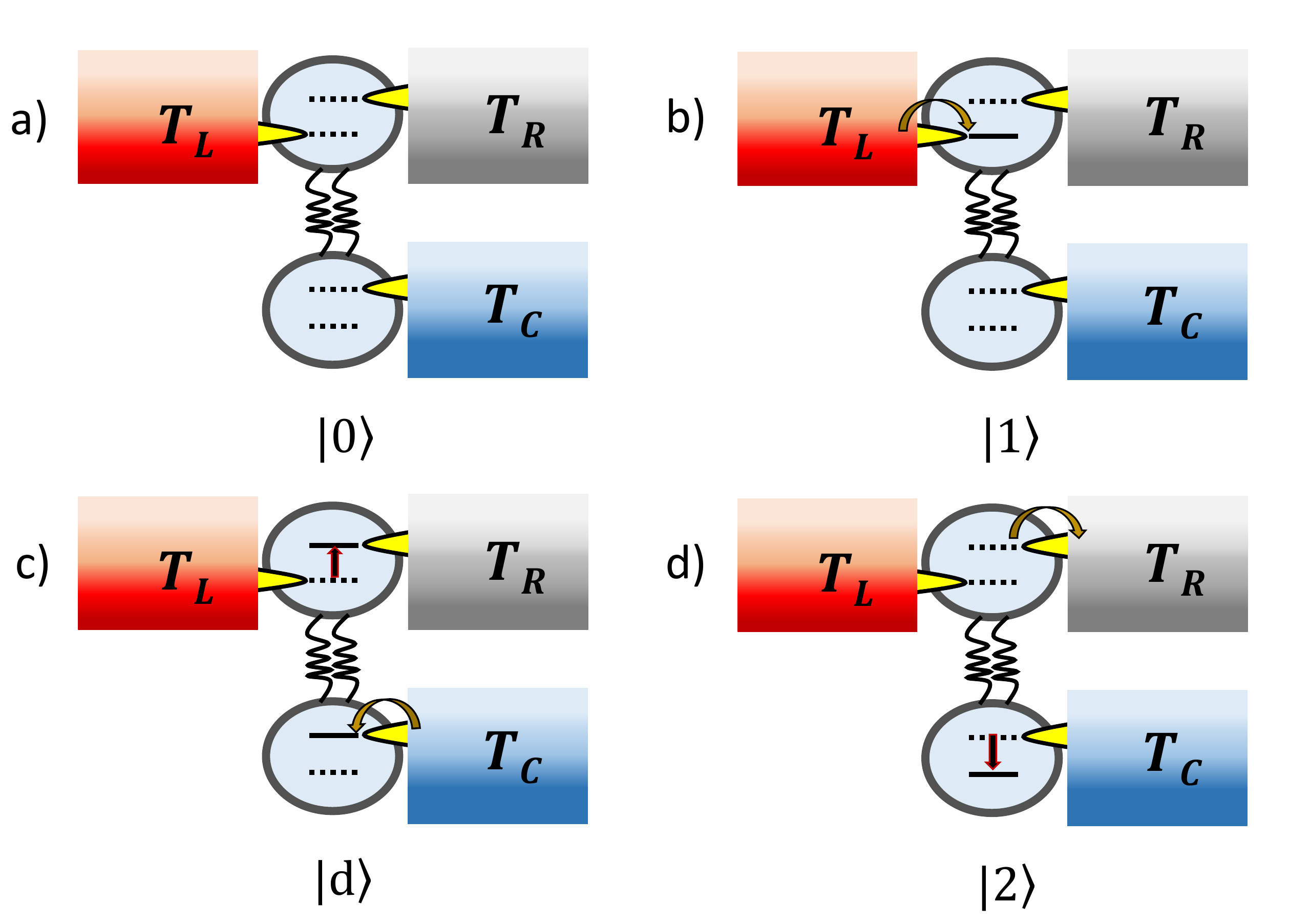}	
	\caption{Cartoon for the cooling mechanism: a) the system is in the ground state, $|0\rangle$, b) Q1 goes to the excited state when a photon enters from the left (L) bath taking the system to state $|1\rangle$, c) Q2 goes to the excited state by absorbing an amount of energy $\epsilon_2 +\Delta_z$ from bath C, d) Q1 goes to the ground by emitting a photon and finally Q2 goes to the ground state by emitting a photon with energy $\epsilon_2$ (not shown in the figure). Once this cycle $|0\rangle \rightarrow |1\rangle \rightarrow |d\rangle \rightarrow |2\rangle \rightarrow |0\rangle$ is completed, the bath C loses an amount of energy equal to $\Delta_z$.}
	\label{fig:mechanism}
\end{figure}

The steady state probabilities for different states can be obtained using the master equation formulation. Once the steady state probabilities are known, calculating heat current flowing out of the bath C is straightforward (see App.~\ref{app:masx0} for details). The heat current flowing out of bath C can be expressed as
\begin{multline}
J_{\rm C}=\epsilon_{2}\left(\Gamma_{02,{\rm C}}p_0-\Gamma_{20,{\rm C}}p_2\right)\\
+(\epsilon_2+\Delta_z)\left(\Gamma_{1d,{\rm C}}p_1-\Gamma_{d1,{\rm C}}p_d\right).
\label{eq:heat_mas}
\end{multline} 
Using the detailed balance equation, along with the probabilities obtained from the master equations, we obtain following condition for cooling
\begin{multline}
e^{\beta_{\rm C}\epsilon_2}\left(e^{\beta_{\rm R}\epsilon_1}\Gamma_{10,{\rm L}}+e^{\beta_{\rm L}\epsilon_1}\Gamma_{10,{\rm R}}\right)\sum\limits_{\alpha={\rm L,R}} e^{\beta_\alpha\epsilon_{d2}}\Gamma_{2d,\alpha}\\
>e^{\sum\limits_{\alpha={\rm L,R}}\beta_\alpha \left(\epsilon_1 +\beta_{\rm C} \epsilon_{d1}\right)}\Gamma_{10}\Gamma_{2d},
\end{multline}
where $\epsilon_{lm}=\epsilon_l-\epsilon_m$ and $\beta_\alpha=[k_{\rm B}T_\alpha]^{-1}$. A similar condition for cooling was obtained in Ref.~\onlinecite{paolo2018} for the case of Coulomb coupled quantum dots. Using the above condition, we observe that $J_{\rm C}$ is a decreasing function of $\Gamma_{10,{\rm R}}$ and $\Gamma_{2d,{\rm L}}$. Choosing the optimal condition, i.e. $\Gamma_{10,{\rm R}}=\Gamma_{2d,{\rm L}}=0$, the condition for cooling reduces to
\begin{equation}
{\beta_{\rm C}\epsilon_2+\beta_{\rm R}\epsilon_{d2}}>{\beta_{\rm C}\epsilon_{d1}+\beta_{\rm L}\epsilon_1}.
\end{equation}
Under the weak coupling approximation, $\Gamma_{10,{\rm R}}=\Gamma_{2d,{\rm L}}\approx 0$ can be obtained by taking a Lorentzian-type spectral density as expressed in Eq.~(\ref{eq:spectral_den_lor}) with $\Omega_{\rm L}=\epsilon_1$ and $\Omega_{\rm R}=\epsilon_1+\Delta_z$, $\Delta_z$ is the energy gain/loss. Taking symmetric thermal bias $T_{\rm L/C}=T\pm \Delta T$, the condition for cooling reduces to a very simple relation given by
\begin{equation}
{\beta_{\rm C}}{\Delta_z}<{\beta_{\rm L}}{\epsilon_1}.
\label{eq:sim_con}
\end{equation}
In terms of Boltzman factor for systems in equilibrium, Eq.~(\ref{eq:sim_con}) becomes $p_{\rm C}^{\rm eq}(\Delta_z)>p_{\rm L}^{\rm eq}(\epsilon_1)$, where $p_{\eta}^{\rm eq}(\epsilon) = e^{-\beta_{\eta}\epsilon}$ is the Boltzman factor for the qubit kept in contact with bath $\eta$ where $\epsilon$ gives the energy gap of the qubit. Hence, refrigeration can be achieved if the Boltzmann factor associated with a qubit of gap $\Delta_z$ and attached to the coldest bath is larger than the Boltzman factor of a qubit with gap $\epsilon_1$ kept in contact with the hottest bath.
\begin{figure}[!htb]
\includegraphics[width=\columnwidth]{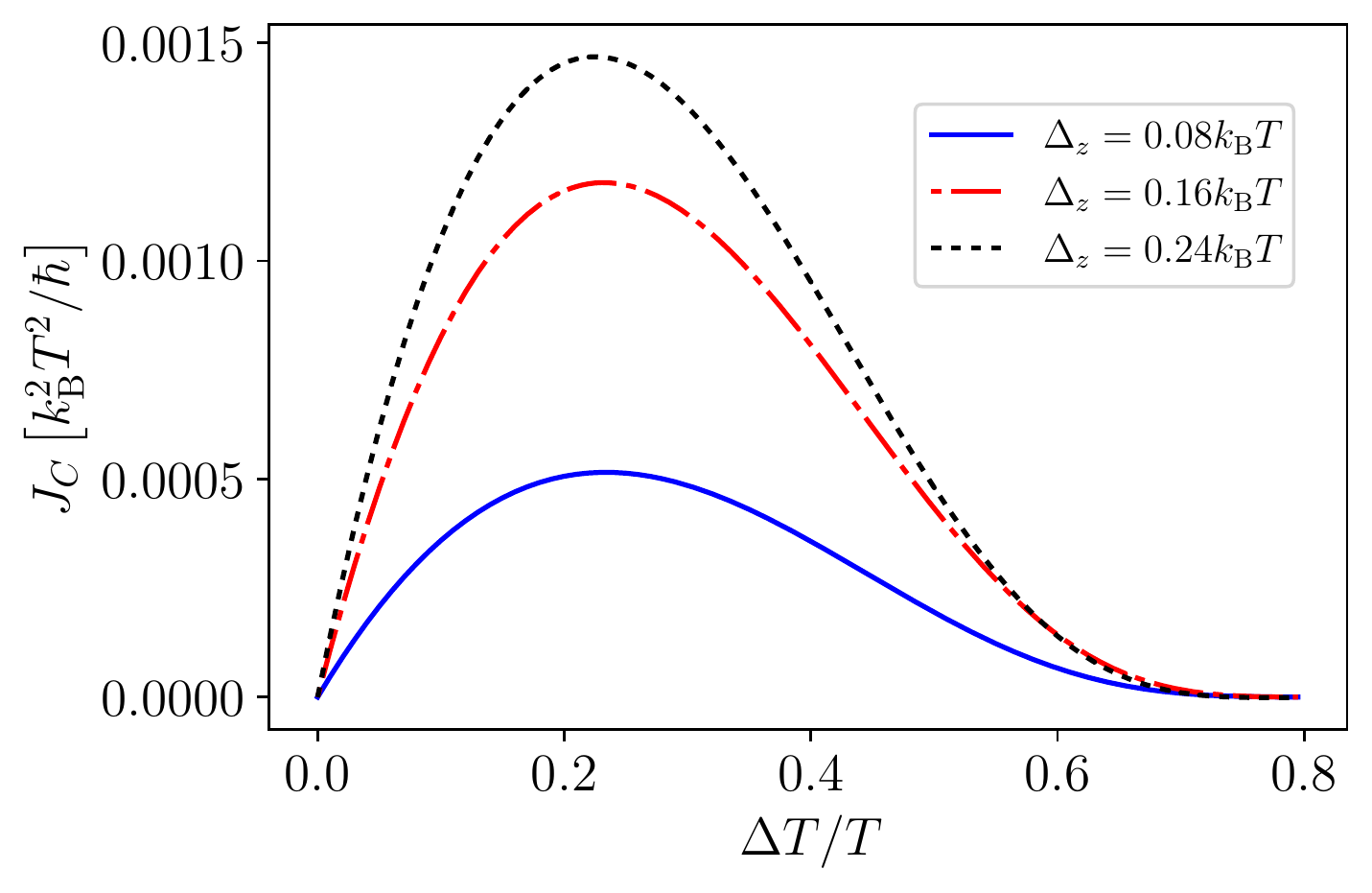}	
	\caption{Heat current flowing out of the bath C as a function of $\Delta T$ using Eq.~(\ref{eq:heat_mas}). Parameters: $\Gamma_{\rm L}=0.02$, $\Gamma_{\rm R}=0.08$, $\Gamma_{\rm C}=0.06$, $\epsilon_1=2 k_{\rm B}T$, $\epsilon_2=2 k_{\rm B}T$, $\Omega_{\rm L}=\epsilon_1$, $\Omega_{\rm R}=\epsilon_1+\Delta_z$, $\Omega_{\rm C}=\epsilon_2+\Delta_z$, $T_{\rm R}=T$, $T_{\rm L}=T+\Delta T$, $T_{\rm C}=T-\Delta T$, $\epsilon_{\rm C}=20 k_{\rm B}T$. }
	\label{fig:currdT}
\end{figure}

In light of the above derived condition for refrigeration, we argue that the process of refrigeration follows the following steps\cite{paolo2018} (see Fig.~\ref{fig:mechanism}): 1) Initially both Q1 and Q2 are in the ground state. Q1 goes to the excite state 1 when a photon enters from one of the baths (L or R) 2) Q2 goes to the excited state absorbing energy $\epsilon_2+\Delta_z$ from the bath C (the system goes to state d) 3) Q1 goes to the ground state by emitting a photon (the system goes to state 2) and 4) Q2 emits a photon with energy $\epsilon_2$ and both qubits are back to the ground state. Each time the cooling cycle is completed, bath C loses an amount of energy equal to $\Delta_z$. There is an alternative cycle which begins with the excitation of Q2 and leads to the dissipation of heat into the cold bath C. We make the cooling cycle dominant by suitably engineering the energy filters. Similar cycles were  considered to study Coulomb drag\cite{sanchez2011} as well as thermal drag \cite{bhandari2018} in Coulomb coupled electronic systems.

In the following, we study the order dependence of the heat current flowing out of bath C as a function of $\Delta_z$. We observe that the heat current $J_{\rm C}$ is second order in $\Delta_z$ and it takes a simple form when expanded over both $\Delta T$ and $\Delta_z$. We obtain {\color{red}from Eq.~(\ref{eq:heat_mas})},
\begin{equation}
J_{\rm C}=\frac{-e^{2\epsilon/k_{\rm B}T} K(\epsilon)K(\epsilon+\Delta_z)(-1+e^{\epsilon/k_{\rm B}T})^{-1}\Delta T}{(1+e^{\epsilon/k_{\rm B}T})^2(K(\epsilon+\Delta_z)+K(\epsilon)e^{\epsilon/k_{\rm B}T})}\frac{\Delta_z^2}{k_{\rm B}^2T^2},
\label{eq:hc_delsq}
\end{equation}
where we assumed $\epsilon_1=\epsilon_2=\epsilon$ and $K_{\rm L}(\epsilon)=K_{\rm R}(\epsilon)
=K_{\rm C}(\epsilon)=K(\epsilon)$.

\begin{figure}[!htb]
\includegraphics[width=\columnwidth]{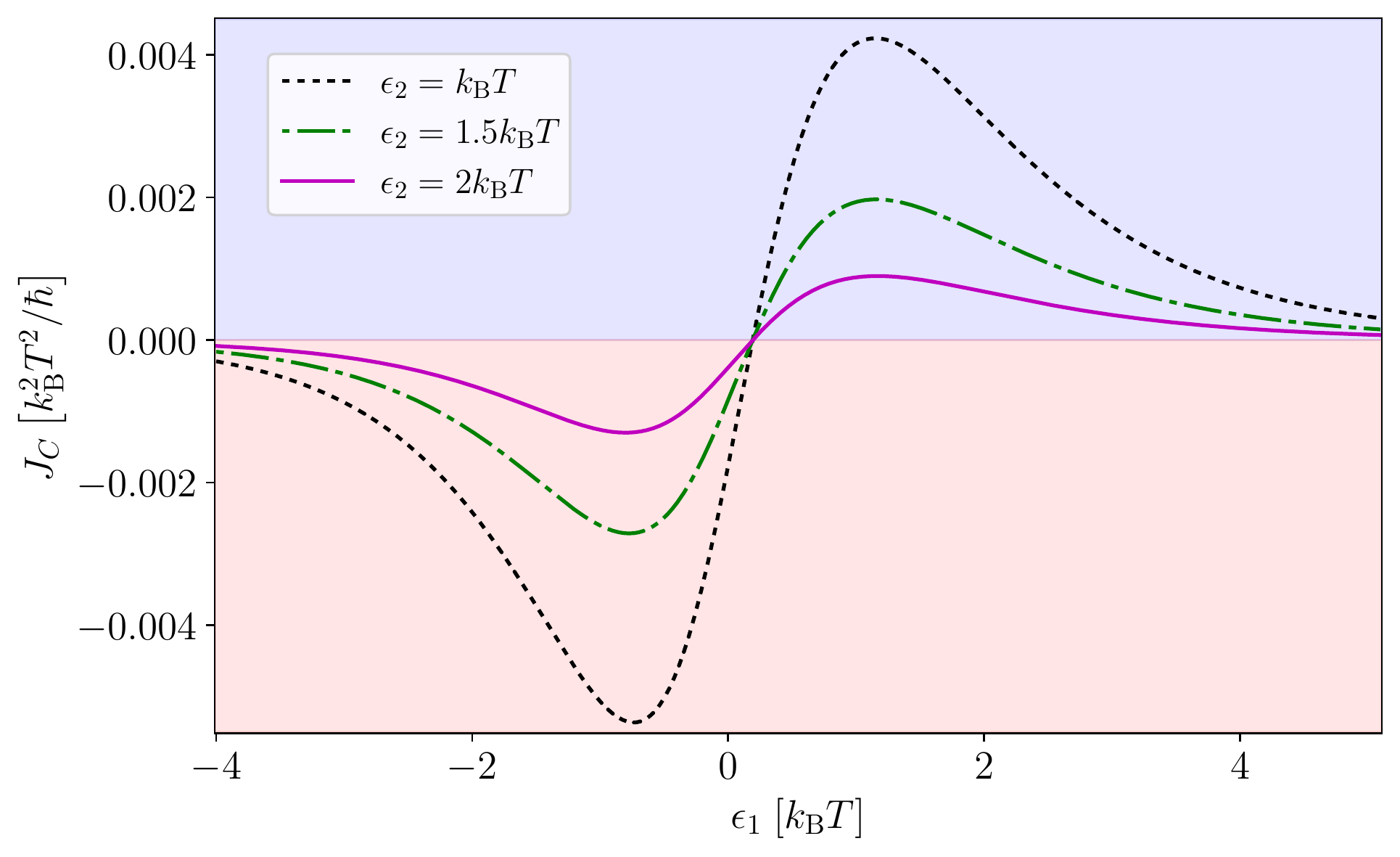}	
	\caption{Heat current flowing into the bath C as a function of $\epsilon_1$ using Eq.~(\ref{eq:heat_mas}). Parameters: $\Delta_z=0.1$, $\Gamma_{\rm L}=0.02$, $\Gamma_{\rm R}=0.08$, $\Gamma_{\rm C}=0.06$, $\Omega_{\rm L}=\epsilon_1$, $\Omega_{\rm R}=\epsilon_1+\Delta_z$, $\Omega_{\rm C}=\epsilon_2+\Delta_z$, $T_{\rm R}=T$, $T_{\rm L}=T+\Delta T$, $T_{\rm C}=T-\Delta T$, $\Delta T=0.3T$, $\epsilon_{\rm C}=10k_{\rm B}T$. In the blue shaded region, the bath C gets cooled whereas in the red shaded region it gets heated.}
	\label{fig:currCe1}
\end{figure}
In Fig.~\ref{fig:currdT}, we plot the heat current flowing out of the bath C as a function of thermal bias $\Delta T$. The system-bath coupling strengths are chosen small enough to be well within the weak coupling regime. The positions of the Lorentzians (in the definition of $K_\alpha(\epsilon)$) are chosen so as to maximize the cooling effect. For the set of parameters considered, we observe cooling in the entire range of thermal bias with a maximum around $\Delta T\approx 0.3 T$. Note that, positive values of $J_{\rm C}$ implies that the heat is flowing out of bath C, i.e. bath C is being cooled. We further observe that the magnitude of cooling depends on the value of $\Delta_z$. When plotted as a function of $\Delta_z$, the heat current shows a maximum at some intermediate values before going to zero for very large values of $\Delta_z$ (not shown in the figure). 

In Fig.~\ref{fig:currCe1}, we plot the heat current as a function of $\epsilon_1$. We show that the heat is dissipated into the bath C (heating) when $\epsilon_1<0 $ whereas for $\epsilon_1>0$ heat gets extracted from bath C (refrigeration). Hence the direction of heat flow in the lower circuit can be tuned by suitably changing the energy of the upper qubit. 
\subsection{General case: Global master equation vs Local Master Equation} 
\label{sec:zzxx}
\begin{figure}[!htb]
\includegraphics[width=\columnwidth]{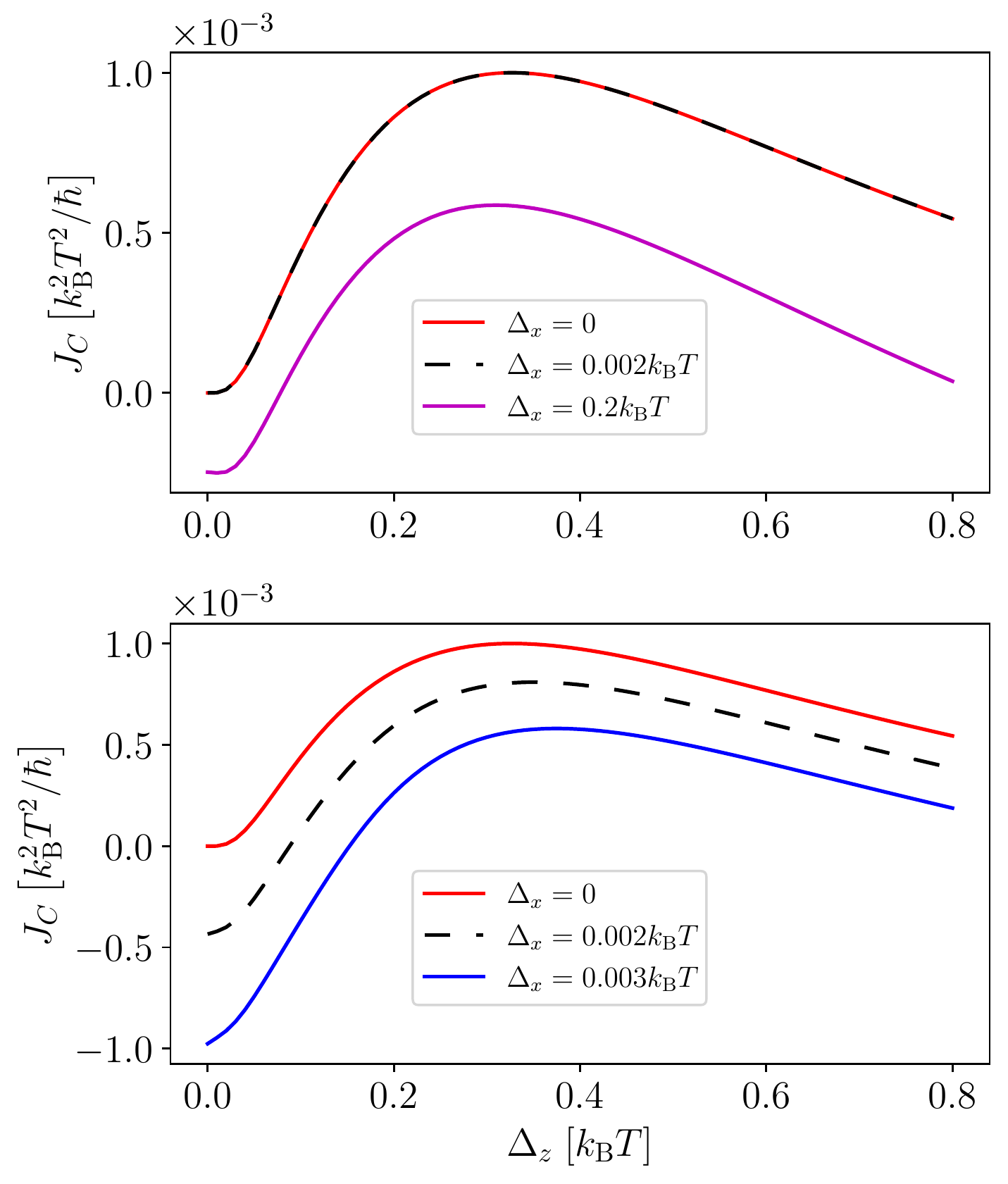}	
	\caption{Heat current flowing out of the bath C as a function of $\Delta_z$. Upper panel is for {\em local master equation} case whereas the lower panel is for {\em global master equation case}. Parameters: $\Gamma_{\rm L}=0.02$, $\Gamma_{\rm R}=0.08$, $\Gamma_{\rm C}=0.06$, $\epsilon_1=2k_{\rm B}T$, $\epsilon_2=2.1k_{\rm B}T$, $\Omega_{\rm L}=\epsilon_1$, $\Omega_{\rm R}=\epsilon_1+\Delta_z$, $\Omega_{\rm C}=\epsilon_2+\Delta_z$, $T_{\rm R}=T$, $T_{\rm L}=T+\Delta T$, $T_{\rm C}=T-\Delta T$, $\Delta T=0.1T$, $\epsilon_{\rm C}=10 k_{\rm B}T$.}
	\label{fig:gloloc}
\end{figure}

In the presence of XX coupling, the derivation of master equation in the diagonalized basis naturally leads to the global master equation whereas local master equation follows from the approach which neglects the presence of XX coupling when computing the effects of the baths. Recently, the range of validity of local and global master equation has received much attention, especially regarding the consistency with the thermodynamic principles in the case of local master equation. It has been observed that the local master equation is a valid approximation only when the XX coupling between the two qubits is sufficiently small. However, global master equation followed by a secular approximation is better suited for intermediate and large values of $\Delta_x$\cite{Cattaneo2019,Hofer2017}.

We study absorption refrigeration using both local and global master equation. In Fig.~\ref{fig:gloloc}, we plot the heat current flowing out of bath C as a function of $\Delta_z$ for a range of values of $\Delta_x$ - in the upper panel we use the local master equation whereas in the lower panel we use the global master equation (see Appendix \ref{app:mas} for details). In both cases, we observe that the XX coupling between the two qubit is detrimental to the cooling effect. However, local master equation predicts that the cooling is not reduced as drastically as predicted by the global master equation for small values of $\Delta_x$. In particular for $\Delta_x=0.002k_{\rm B}T$ (black dashed curves in both panels), global master equation predicts heating of the cold bath C (note that negative heat current means positive heat flowing into the bath) for $\Delta_z\lesssim 0.2 k_{\rm B}T$ whereas the local master equation gives cooling effect in that parameter regime. As mentioned previously, in the regime of interest, i.e. for small values of $\Delta_x$, local master equation is better suited to describe the dynamics.  
\section{Absorption refrigerator: two coupled non linear resonators}
\label{sec:res}
\begin{figure}[!htb]
\includegraphics[width=\columnwidth]{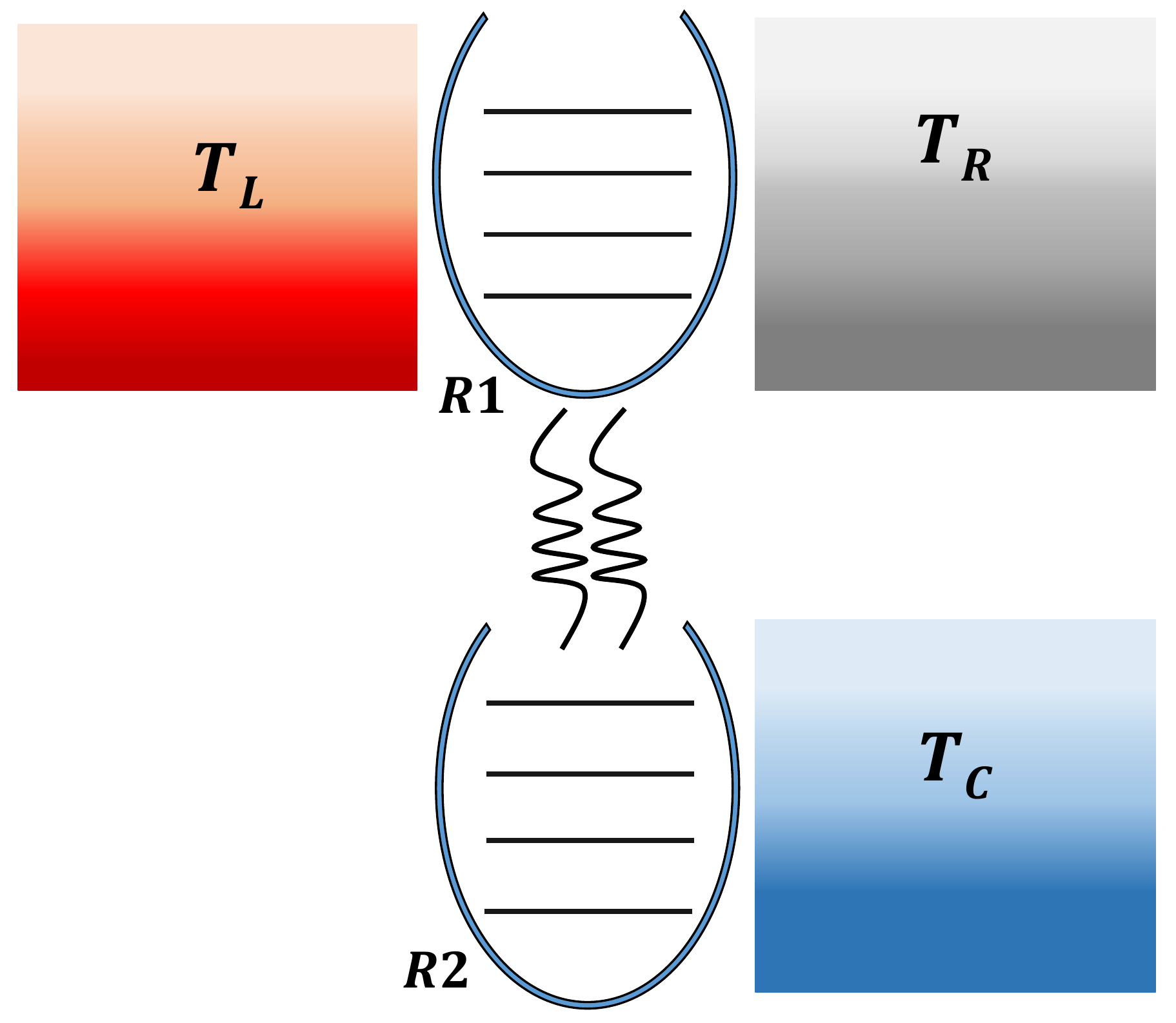}	
	\caption{The model for absorption refrigerator based on two non-linearly coupled resonators}
	\label{fig:setup3}
\end{figure} 
In this section, we investigate absorption refrigeration in non-linearly coupled resonators. As shown in Fig.~\ref{fig:setup3}, the setup consists of two resonators R1 and R2 non-linearly coupled to each other. As in the previous section, the upper resonator R1 is attached to two baths - one of them is kept at relatively higher temperature whereas the lower resonator R2 is attached to the coldest bath. The system Hamiltonian is \begin{equation}
H_S=\epsilon_1 a_1^\dagger a_1 +\epsilon_2 a_2^\dagger a_2+\Delta_z a_1^\dagger a_1 a_2^\dagger a_2,
\end{equation}
where $\epsilon_i/\hbar$ is the frequency of the resonator $i$ and $\Delta_z$ is the parameter determining the strength of non-linearity. We consider bosonic baths as defined in Eq.~(\ref{eq:ham_bath}) and the coupling Hamiltonian takes the following form
\begin{equation}
H_C=\sum_{k,\alpha={\rm L,R}}V_{k\alpha}\,a_1^\dagger b_{k\alpha}+\sum_k V_{kC}\,a_2^\dagger b_{kC}+h.c.,
\end{equation}
where we neglected the counter rotating terms. The counter rotating terms do not give significant effects in the weak coupling regime. In this section, we will study heat transport using two different formulations: 1) master equation technique and 2) Keldysh nonequilibrium Green's function formalism.

\subsection{Master Equation Calculations}
\label{sec:resmas}
We start our analysis of coupled resonator system from the weak system-bath coupling regime. In this regime, the dynamics can be studied perturbatively in terms of system-bath coupling $H_C$. Up to the leading order sequential contribution (Born approximation), Linblad formulation can be used to obtain the master equation for the reduced system density matrix, provided dynamics of the baths is fast compared to the system dynamics (Markov approximation). The heat transport is carried out by sequential tunneling processes where all possible single photon processes (both incoming (absorption) and outgoing (emission)) are taken into account. The details of the calculation is presented in the App.~\ref{app:masres}. The master equation in the photon number basis reads
\begin{widetext}
\begin{multline}
\dot{\rho}_{n_1,n_2}=-\bigg[\Big(C_1({n_1},{n_2})+D_1({n_1},{n_2})+C_2({n_1},{n_2})+D_2({n_1},{n_2})\Big)\rho_{n_1,n_2}-C_1({n_1-1},{n_2})\rho_{n_1-1,n_2}\\
-D_1({n_1+1},{n_2})\rho_{n_1+1,n_2}-C_2(n_1,{n_2-1})\rho_{n_1,n_2-1}-D_2(n_1,{n_2+1})\rho_{n_1,n_2+1}\bigg],
\end{multline}
\end{widetext}
where the incoming transition rates that brings the system into the final state $({n_1+1,n_2})$ and $({n_1,n_2+1})$ from the state $({n_1,n_2})$ are respectively given by
\begin{align}
&C_1({n_1},{n_2})=2\sum_{\alpha={\rm L,R}}(n_1+1){\cal }F_\alpha(\tilde
{\omega}_{1,n_2}),
\nonumber\\
&C_2({n_1},{n_2})=2(n_2+1){\cal F}_{\rm C}(\tilde
{\omega}_{2,n_1})
\end{align}
and the outgoing transition rates that leads the system out of the state $({n_1,n_2})$ are given by
\begin{align}
&D_1({n_1},{n_2})=\sum_{\alpha={\rm L,R}}2n_1{\cal G}_\alpha(\tilde
{\omega}_{1,n_2}),\nonumber\\
&D_2({n_1},{n_2})=2n_2{\cal G}_{\rm C}(\tilde
{\omega}_{2,n_1}),
\end{align}
where $\tilde{\omega}_{1,n_2}=\epsilon_1+\Delta_z n_2$, $\tilde{\omega}_{2,n_1}=\epsilon_2+\Delta_z n_1$ and
\begin{align}
&{\cal F}_\eta({\tilde{\omega}})=\frac{1}{2} K_\eta({\tilde{\omega}})
n_\eta({\tilde{\omega}})\nonumber \\
&{\cal G}_\eta({\tilde{\omega}})=\frac{1}{2} K_\eta({\tilde{\omega}})
(1+n_\eta
({\tilde{\omega}})),  
\end{align} 
where $n_\eta(\omega)$ is the Bose-Einstein distribution for bath $\eta$. Similarly, the heat current flowing out of the bath C can be written as
\begin{equation}
J_{\rm C}={\rm Tr}\left[H_S{\cal L}_{\rm C}\rho\right],
\label{eq:cur_trace}
\end{equation}
where ${\cal L}_{\rm C}$ is the part of Lindbladian associated with the bath C given by
\begin{multline}
[{\cal L}_{\rm C} \rho]_{n_1,n_2}=-\Big(C_2({n_1},{n_2})
+D_2({n_1},{n_2})\Big)\rho_{n_1,n_2}\\
+C_2(n_1,{n_2-1})\rho_{n_1,n_2-1}+D_2(n_1,{n_2+1})\rho_{n_1,n_2+1}.
\end{multline}
\begin{figure}[!htb]
\includegraphics[width=\columnwidth]{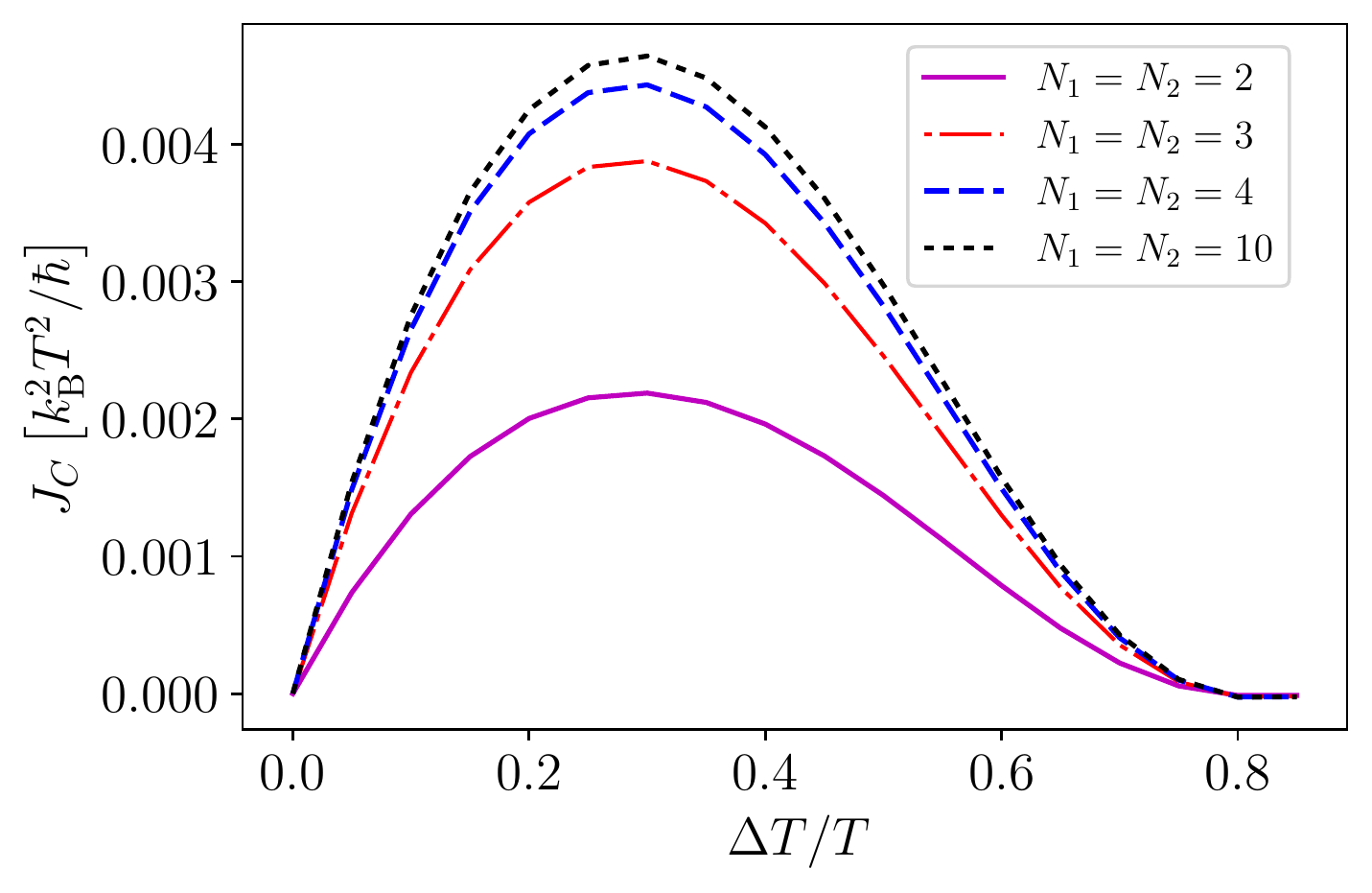}	
	\caption{Heat current flowing out of the bath C as a function of thermal bias, $\Delta T$. Parameters: $\Delta_z =0.05 k_{\rm B}T$, $\Gamma_{\rm L}=0.02$, $\Gamma_{\rm R}=0.08$, $\Gamma_{\rm C}=0.06$, $\epsilon_2=\epsilon_1=k_{\rm B}T$, $\Omega_{\rm L}=\epsilon_1$, $\Omega_{\rm R}=\epsilon_1+\Delta_z$, $\Omega_{\rm C}=\epsilon_2+\Delta_z$, $T_{\rm R}=T$, $T_{\rm L}=T+\Delta T$, $T_{\rm C}=T-\Delta T$, $\epsilon_{\rm C}=10k_{\rm B}T$. $N_1$ and $N_2$ give the number of levels considered for R1 and R2 respectively for the numerical calculations.}
	\label{fig:res_qbt_dT}
\end{figure}

Although the number of levels in each resonator ranges to infinity, under the low temperature and weak coupling approximation, only the few lower levels participate in transport. We define $N_i$ as the the number of levels we consider for the resonator $i$ while undergoing the trace in Eq.~(\ref{eq:cur_trace}). Hence, $N_i=\text{max}(n_i)$, for $i=1,2$. For instance, in the limit of $N_1=N_2=2$ the coupled resonator system reduces to the coupled qubit system with ZZ interaction studied in Sec.~\ref{sec:zz}. In Fig.~\ref{fig:res_qbt_dT}, we study the effect of thermal bias on the heat current $J_{\rm C}$. We compare the cooling effect for different $N_1,N_2$. In the parameter regime considered, we observe that the multi-level systems ($N_{1},N_2>2$) provide a more stronger cooling effect compared to the coupled qubit system ($N_{1},N_2=2$). For $\Delta T\approx 0.3 T$, cooling effect is more than double in the resonator case (black dashed curve) compared to the qubits (magenta curve). Reasonably, the largest correction beyond the qubit-limit comes from the qutrit case ($N_1,N_2=3$). However, the correction beyond the qutrit limit are significant, although small.
\begin{figure}[!htb]
\includegraphics[width=\columnwidth]{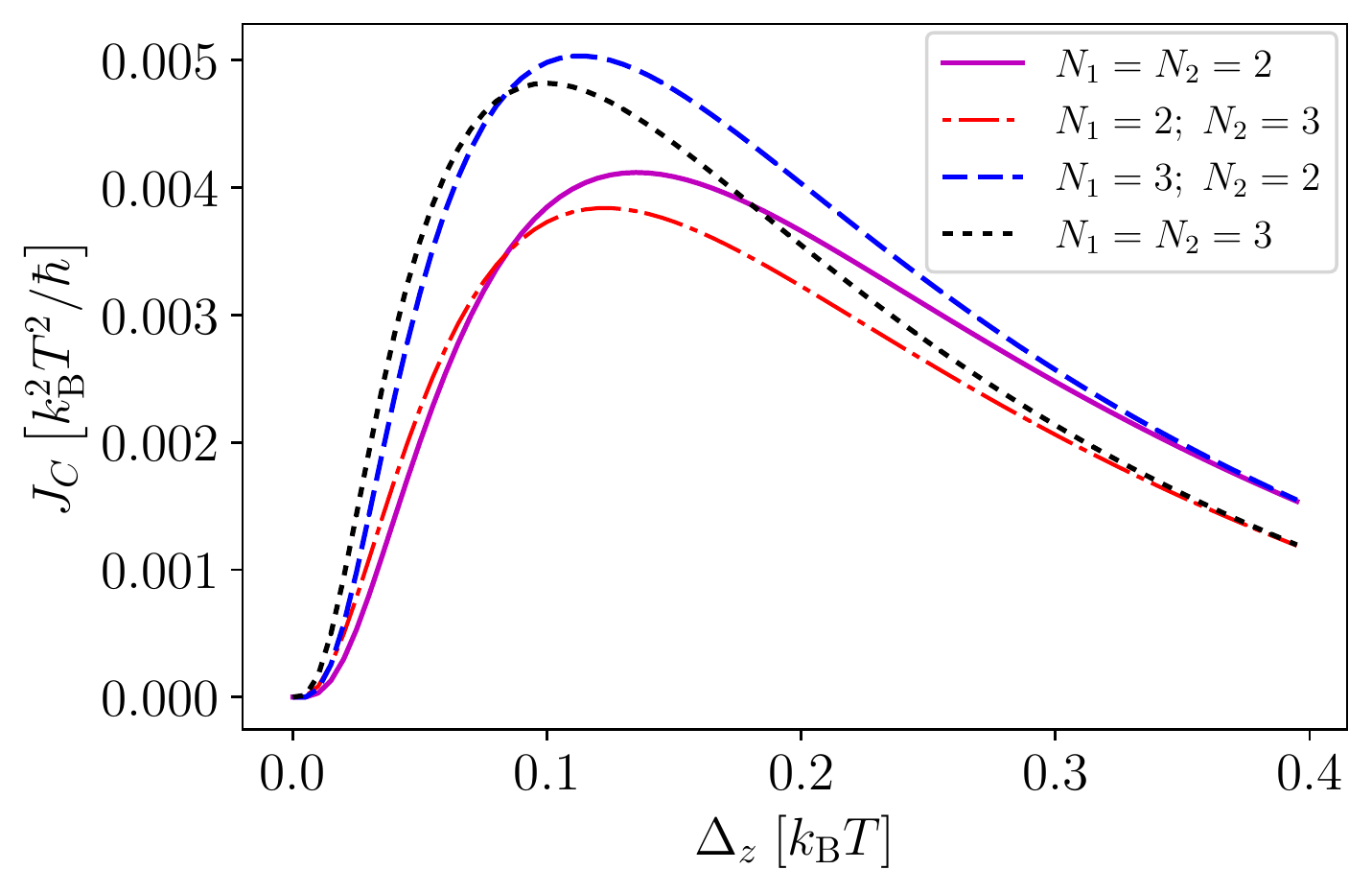}	
	\caption{Heat current flowing out of the bath C as a function of $\Delta_z$. Parameters: $\Gamma_{\rm L}=0.02$, $\Gamma_{\rm R}=0.08$, $\Gamma_{\rm C}=0.06$, $\epsilon_2=\epsilon_1=k_{\rm B}T$, $\Omega_{\rm L}=\epsilon_1$, $\Omega_{\rm R}=\epsilon_1+\Delta_z$, $\Omega_{\rm C}=\epsilon_2+\Delta_z$, $T_{\rm R}=T$, $T_{\rm L}=T+\Delta T$, $T_{\rm C}=T-\Delta T$, $\Delta T=0.3T$, $\epsilon_{\rm C}=20$. $N_1$ and $N_2$ give the number of levels considered for R1 and R2 respectively for the numerical calculations.}
	\label{fig:res_qbt_del}
\end{figure}
	 
As observed in Fig.~\ref{fig:res_qbt_dT}, the major contribution to the heat current $J_{\rm C}$ comes from the first three levels of each resonator.
In Fig.~\ref{fig:res_qbt_del}, we study heat current flowing out of bath C as a function of $\Delta_z$ for $2\leq N_1,N_2\leq 3$. In all the cases, we observe that heat current reaches a maximum before decreasing monotonously. We will explain this behavior taking the qubit-qubit case, i.e. $N_1=N_2=2$ (magenta curve). As discussed in Sec.~\ref{sec:zz}, cooling takes place when a photon leaves bath C with an energy $\epsilon+\Delta_z$ and enters with $\epsilon$. Although for small values of $\Delta_z$ (close to $0$), there are many photons that participate in transport, the amount of heat they carry out of bath C is significantly small. With increasing $\Delta_z$, the amount of energy carried out of the bath C increases but at the cost of number of photons. For large enough values of $\Delta_z$, the thermal energy of bath C becomes insufficient to provide enough photons that could possibly excite the resonator, which in turn leads to a decrease of heat current. Moreover, for large values of $\Delta_z$, the state where both resonators are excited becomes less probable which is crucial in achieving the cooling effect. Similar reasoning can be given for the multi-level case. In addition, we observe that for $\Delta_z\lesssim 0.2 k_{\rm B}T$, cooling is enhanced for $N_1=3$. This is because, for small values of $\Delta_z$ the resonator R2 can get excited either to the energy state $\epsilon_2+\Delta_z$ or $\epsilon_2+2\Delta_z$ depending on whether R1 is in the energy state $\epsilon_1$ or $2\epsilon_1$ respectively. This creates an extra channel for the refrigeration process giving enhanced cooling effect. However for $\Delta_z>0.35 k_{\rm B}T$, the heat current $J_{\rm C}$ becomes independent of $N_1$ as long as $N_1\geq 2$. For large values of $\Delta_z$, the excitation of R2 to states $\epsilon_2+2\Delta_z$ and $2\epsilon_2+2\Delta_z$ becomes less probable. In addition, these processes become energetically costly. Hence, only the first two levels of R1 determine the transfer of heat in the bath C.
\subsection{Non equilibrium Green's function calculations }
\label{sec:resgreen}
In this section we will address the effect of higher order terms corresponding to system-bath coupling on absorption refrigeration. The amount of energy absorbed or emitted under a first order sequential tunneling process is given by the difference of energy of initial and final states. A photon must have this amount of energy to excite the qubit. This creates an ideal situation for engineering energy filters. However, higher order processes such as cotunneling can occur via virtual states even if the photons do not have enough energy to excite the qubit. These processes can destroy the filtering effect which are designed taking only the sequential processes into account. In this section, we will employ the Keldysh nonequilibrium Green's function to address the aforementioned scenario. We define the retarded Green's function for the system as
\begin{equation}
G_{i;j}^r(t,t')=-i\theta(t-t')\left\langle\left[a_i(t),a_i^\dagger(t')\right]\right\rangle.
\end{equation}
We use the equation of motion to obtain following Dyson equation for R2 in the mean field approximation 
\begin{multline}
i\partial_t G_{2;2}^r(t,t')=\delta(t-t')+\epsilon_2 G_{2;2}^r(t,t')+\Delta_z \left\langle n_1\right\rangle G_{ 2;2}^r(t,t')\\
+\int dt_1 \Sigma_{\rm C}^{r}(t,t_1)G_{2;2}^r(t_1,t'),
\end{multline}
 $n_1=a_1^\dagger a_1$ and the self energy due to coupling to bath C is given by
\begin{equation}
\Sigma_{\rm C}^r(t_1,t_2)=\sum_k |V_{kC}|^2 g_{kC}^r(t_1,t_2),
\end{equation}
where $g_{kC}^r(t_1,t_2)=-i\theta(t_1-t_2)\left\langle\left[b_{kC}(t_1),b_{kC}^\dagger(t_2)\right]\right\rangle$ is the retarded Green's function for the free bath.
The heat current flowing out of the bath C at time $t$ can be written as
\begin{equation}
J_{\rm C}(t)=\frac{d}{dt}\left\langle H_{B,{\rm C}}(t)\right\rangle
\end{equation}
Using the Meir Wingreen approach, the heat current in the steady state takes the final form given by\cite{meir1992}
\begin{equation}
J_{\rm C}=\int \frac{d\epsilon}{2\pi}\epsilon \left[G_{22}^>(\epsilon)\Sigma_{\rm C}^<(\epsilon)-G_{22}^<(\epsilon)\Sigma_{\rm C}^>(\epsilon)\right],
\label{eq:curr_negf}
\end{equation}
where the lesser component of self energy, $\Sigma_\eta^<=-in_
\eta K_\eta(\epsilon)$ and the greater component, $\Sigma_\eta^>(\epsilon)=-iK_\eta(\epsilon) \left(1+n_\eta(\epsilon)\right)$. The existence of steady state in interacting open quantum systems has been proved in Refs.~\onlinecite{referee1,referee2}.  $G_{ij}^>(\epsilon)$ and $G_{ij}^<(\epsilon)$ are the Fourier transform of the greater and lesser Green's functions respectively defined through the relations
\begin{align}
G_{ij}^>(t,t')&=-i\left\langle a_i(t)a_j^\dagger(t')\right\rangle\nonumber \\
G_{ij}^<(t,t')&=-i\left\langle a_j^\dagger(t')a_i(t)\right\rangle.
\end{align} 
They can be calculated using the following relation
\begin{equation}
G_{2;2}^{\lessgtr}(\epsilon)={G}_{2;2}^{r}(\epsilon){\Sigma}_{\rm tot,2}^{\lessgtr}(\epsilon){G}_{2;2}^{a}(\epsilon),
\end{equation}
where the self-energy
\begin{equation}
\Sigma_{\rm tot,2}^{\lessgtr}=\Sigma_{\rm C}^{\lessgtr}+\Sigma_{int,2}^{\lessgtr},
\label{eq:tot_self}
\end{equation}
\begin{figure}[!htb]
\includegraphics[width=0.5\columnwidth]{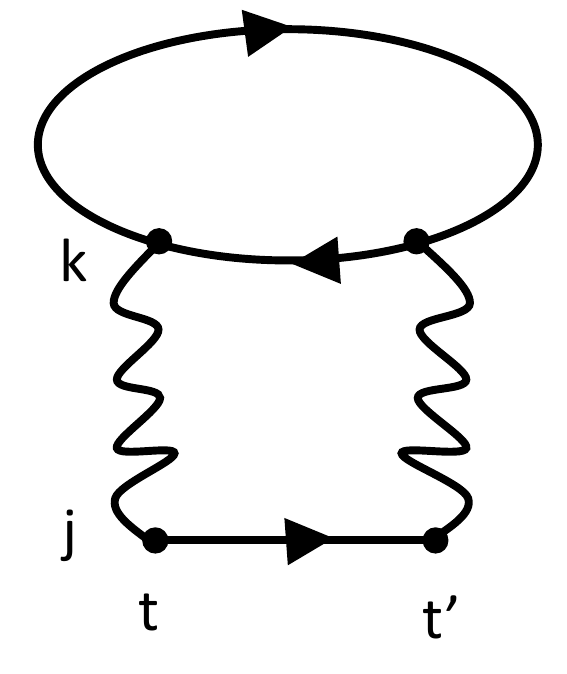}	
	\caption{Single bubble Feynman diagram for the exchange correlation where $j,k=1,2$ depending on the resonator. Note that, $j\neq k$.}
	\label{fig:xc}
\end{figure} 
The first term on the right hand side of Eq.~(\ref{eq:tot_self}) corresponds to the self-energy due to tunneling whereas the second term results from the inter-resonator non-linearity\cite{sanchez}. The latter self-energy can be broken down into two parts $\Sigma_{int,i}=\Sigma_{H,i}+\Sigma_{xc,i}$, where the first term is the Hartree contribution and the second term is the exchange correlation part. Note that $\Sigma_{H,i}^{\lessgtr}=0$. Hence,
\begin{equation}
G_{2;2}^{\lessgtr}(\epsilon)={G}_{2;2}^{r}(\epsilon)\Big(\Sigma_{\rm C}^{\lessgtr}(\epsilon)+\Sigma_{xc,2}^{\lessgtr}(\epsilon)\Big)(\epsilon){G}_{2;2}^{a}(\epsilon).
\label{eq:keldysh_rel}
\end{equation}
The exchange correlation self-energy can be expressed as\cite{stefanucci,sanchez,moldoveanu2009}
\begin{equation}
\Sigma_{j,xc}^{\lessgtr}(t,t')=-\Delta_z^2 G_{j;j}^{\lessgtr}(t,t')P_k^{\lessgtr}(t,t'),
\end{equation}
where 
$P_k^{\lessgtr}(t,t')=G_{k;k}^{\lessgtr}(t,t')G_{k;k}^{\gtrless}(t',t)$\cite{sanchez}. We use mean field Green's function to calculate the exchange correlation self energy. Note that $\left\langle n_j\right\rangle= i\int d\epsilon/2\pi G_{j;j}^<(\epsilon)$ enters in the expression for retarded Green's function. We would solve the retarded and the lesser Green's function self-consistently.
\begin{figure}[!htb]
\includegraphics[width=\columnwidth]{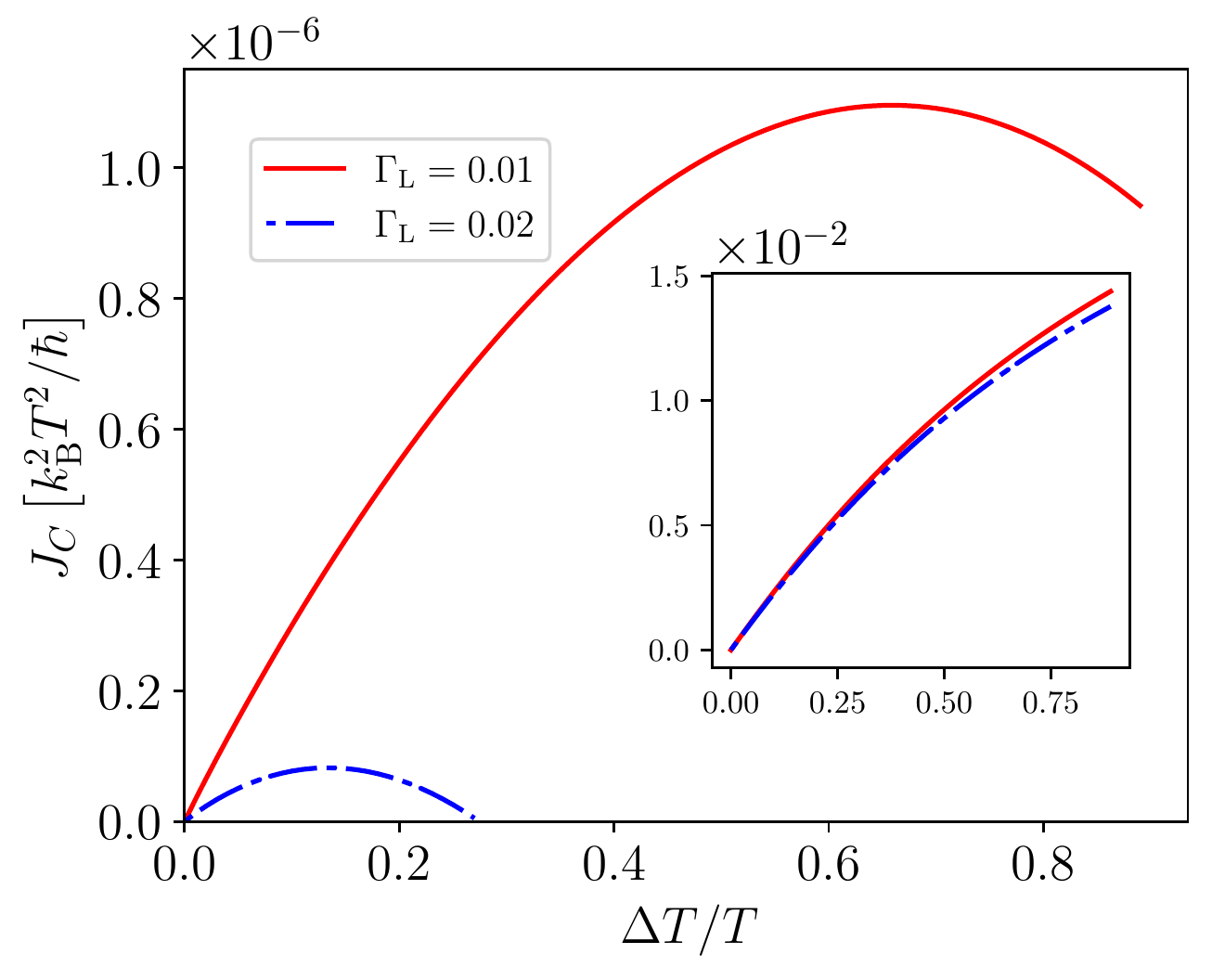}	
	\caption{Heat current flowing out of the bath C as a function of $\Delta T$ for different values of $\Gamma_{\rm L}$ using the nonequilibrium Green's function calculations. In the inset, we present the results obtained using the master equation calculations. Parameters: $\Gamma_{\rm R}=\Gamma_{\rm C}=0.02$, $\epsilon_2=2k_{\rm B}T$, $\epsilon_1=2k_{\rm B}T$, $\Omega_{\rm L}=\epsilon_1$, $\Omega_{\rm R}=\epsilon_1+\Delta_z$, $\Omega_{\rm C}=\epsilon_2+\Delta_z$, $\Delta_z =0.2 k_{\rm B}T$, $\epsilon_{\rm C}=7k_{\rm B}T$, $T_{\rm R}=T$, $T_{\rm L}=T+\Delta T$ and $T_{\rm C}=T$}
	\label{fig:JC_res_mf}
\end{figure} 
If we neglect the exchange correlation part of the self energy, the heat current in Eq.~(\ref{eq:curr_negf}) using Eq.~(\ref{eq:keldysh_rel}) goes to zero. The only non-zero contribution to the heat current $J_{\rm C}$ comes from the exchange correlation part of the self energy. We obtain
\begin{multline}
J_{\rm C}=\int \frac{d\epsilon}{2\pi}\epsilon G_{2;2}^r(\epsilon)G_{2;2}^a(\epsilon)\Big[\Sigma_{xc,2}^>(\epsilon)\Sigma_{\rm C}^<(\epsilon)\\
-\Sigma_{xc,2}^<(\epsilon)\Sigma_{\rm C}^>(\epsilon)\big]
\end{multline}
Note that the mean field approximation is obtained by decoupling the equation of motion for the time-ordered Green's function at first order (see Appendix \ref{app:greenres} for details). Although the mean field approximation addresses the strong system-bath coupling going beyond the sequential tunneling processes, it neglects tunneling processes resulting from strong non-linearity. These processes may contribute significantly to the refrigeration process for strong enough non-linearity. Note that, our calculations hold for small to intermediate strength of non-linearity\cite{sanchez}. In order to properly address both strong system-bath coupling and strong non-linearity, one would have to go beyond the mean-field Hartree approximation\cite{rectbibek} which is beyond the scope of present work. In this section, the temperature of the left and the cold bath are taken as $T_{\rm L}=T+\Delta T$ and $T_{\rm C}=T$ respectively. 

We observe no cooling effect when the non-linearity is considered very weak, $\Delta_z\ll\pi \Gamma_\eta \epsilon_{1/2}$ for $\epsilon_{1/2}=2k_{\rm B}T$ in contrast to the master equation calculations (see Fig.~\ref{fig:res_qbt_del}).  The presence of cotunneling and higher order processes in the nonequilibrium Green's function calculations destroys the cooling effect in the weakly non-linear regime. This further demonstrates the importance of higher order terms such as cotunneling  which are often neglected in the literature. In Fig.~\ref{fig:JC_res_mf}, we plot the heat current $J_{\rm C}$ as a function of $\Delta T$ for a couple of values of $\Gamma_{\rm L}$. We consider the strength of non-linearity to be small enough such that $\Delta_z = 0.2 k_{\rm B}T\ll \epsilon_{1/2},k_{\rm B}T$. We observe a large decrease in cooling effect when the coupling strength in the left contact is increased from $\Gamma_{\rm L}=0.01$ to $\Gamma_{\rm R}=0.02$ which is not the case for the master equation calculation (see the inset). In addition, the cooling obtained from the master equation calculations is much larger compared to the nonequilibrium Green's function calculations. The reduction of cooling can be due to two different factors: presence of cotunneling and higher order process, and the mean field approximation which neglects the tunneling processes resulting from strong non-linearity. The heat current approaches a maximum at $\Delta T\approx 0.15 T$ for $\Gamma_{\rm L}=0.02$ and at $\Delta T\approx 0.7 T$ for $\Gamma_{\rm L}=0.01$ before decreasing monotonously for larger values of $\Delta T/T$. The observation of finite cooling effect proves the effectiveness of the proposed models for absorption refrigerators.

\section{Conclusions}
We presented a minimal model for two-body absorption refrigerator based on two qubits coupled via ZZ coupling. We derived the necessary conditions for cooling and also introduced the cooling mechanism. We showed that other isotropic couplings, namely XX and YY coupling can't produce cooling effect. Instead, they are detrimental to cooling. We verified above argument considering both local as well as global master equations. In Sec.~\ref{sec:res}, we studied absorption refrigeration in non-linearly coupled resonators. We compared the magnitude of refrigeration in the mentioned two setups. Non-linearly coupled resonators (multi-level system) was observed to produce better cooling effects in certain parameter regime. We also studied the process of refrigeration in the case of resonators using the Keldysh non-equilibrium Green's function formalism which holds for small to intermediate values of non-linearity. Particularly, we considered the mean field Hartree approximation for the Green's functions and took exchange correlations into account for the self energy.  Albeit small we observed finite cooling effect using the nonequilibrium Green's function formulation further validating the effectiveness of our model as absorption refrigerator. There can be basically three reasons for such a reduction in cooling effect: 1) the higher order processes in terms of system-bath coupling broadens the energy window for transport, hence reducing the cooling effect 2) the higher order processes in terms of non-linearity are not properly addressed which might have an impact on cooling and 3) even a small broadening of energy window can lead to drastic changes in the flow of heat current in the lower circuit which is non-locally coupled to upper circuit. However, the most interesting result is that although we observe cooling effect even for small values of non-linearity with master equation calculations, the non-equilibrium Green's function calculations gives no cooling effect for small $\Delta_z$. The non-equilibrium Green's function calculations are exact for small values of non-linearity.  

To conclude, we find that the best regime to obtain absorption refrigeration in multi-level quantum systems is the weak system-bath coupling regime with intermediate to strong non-linearity.

\begin{acknowledgments}
We thank Paolo Andrea Erdman for insightful discussions and for his comments on the draft. This work was supported by the U.S. Department of Energy (DOE), Office of Science, Basic Energy Sciences (BES), under Award No. DE-SC0017890.
\end{acknowledgments}

\appendix
\section{Master equation for coupled qubit system}
\label{app:mas}
\subsection{Global master equation}
\label{app:masglob}
In order to derive the ``global master equations'' we diagonalise the system Hamiltonian in Eq.~(\ref{eq:sysqbt}), such that
\begin{equation}
H_S=\begin{bmatrix}
E_0 & 0 & 0 &0 \\
0 & E_+ & 0 &0 \\
0 & 0 & E_- &0 \\
0 & 0 & 0 & E_d 
\end{bmatrix},
\end{equation}
where $E_{0/d}=\epsilon_{0/d}$ and 
\begin{equation}
E_{+-}=\frac{\epsilon_1+\epsilon_2}{2}\pm \frac{1}{2}\sqrt{\left(\epsilon_1-\epsilon_2\right)^2+4
\Delta_x^2}.
\end{equation}
The contact Hamiltonian gets modified to
\begin{multline}
H_C=\sum_{k,\eta}\sum_{l=\pm}V_{k\eta}\left(\lambda_{\eta,0l}|0\rangle\langle l|+\lambda_{\eta,l0}|l\rangle\langle 0|\right)\left(b_{k\eta}+b_{k\eta}^\dagger\right)\\
+\sum_{k,\eta}\sum_{l=\pm}V_{k\eta}\left(\lambda_{\eta,dl}|d\rangle\langle l|+\lambda_{\eta,ld}|l\rangle\langle d|\right)\left(b_{k\eta}+b_{k\eta}^\dagger\right),
\end{multline}
where $\eta={\rm L,R,C}$. We have
\begin{align}
&\lambda_{{\rm L},0+}=\lambda_{{\rm R},0+}=-\lambda_{{\rm C},0-}=\cos{\theta}/2,\\
&\lambda_{{\rm L},0-}=\lambda_{{\rm R},0-}=\lambda_{{\rm C},0+}=\sin{\theta}/2,\\
&\lambda_{{\rm L},d+}=\lambda_{{\rm R},d+}=\lambda_{{\rm C},d-}=\sin{\theta}/2,\\
&\lambda_{{\rm L},d-}=\lambda_{{\rm R},d-}=-\lambda_{{\rm 
C},d+}=-\cos{\theta}/2,
\end{align} 
where $\theta = tan^{-1}\left[2\Delta_x/\left(\epsilon_1-\epsilon_2\right)\right]$ and $\lambda_{\eta,lm}=\lambda_{\eta,ml}$. Following the calculations of Ref.~\onlinecite{bibekgreen}, we obtain following master equation
\begin{equation}
\frac{d\mathbf{p}}{dt}=\mathbf{W}\cdot \mathbf{p},
\end{equation}
where
\begin{equation}
\mathbf{p}=\begin{bmatrix}
&p_0 \\
&p_+ \\
&p_- \\
&p_d
\end{bmatrix},
\end{equation}
and 
\begin{multline}
\mathbf{W}=\\\begin{bmatrix}
&-\sum\limits_{n=+,-} \Gamma_{0n} & 
\Gamma_{+0} & \Gamma_{-0} & 0\\
&\Gamma_{0+} & -\sum\limits_{n=0,d} \Gamma_{+n} & 0 & \Gamma_{d+}\\
&\Gamma_{0-} & 0 & -\sum\limits_{n=0,d} \Gamma_{-n} & \Gamma_{d-}\\
&0 & \Gamma_{+d} & \Gamma_{-d} & -\sum\limits_{n=+,-}\Gamma_{dn}
\end{bmatrix}.
\end{multline}
Different transition rates are given by
\begin{equation}
\Gamma_{mn}=\sum_\eta \lambda_{\eta,mn}^2\left(\tilde{\gamma}_\eta(E_{mn})+\gamma_\eta (E_{nm})\right),
\label{eq:tra_rate}
\end{equation}
where $\gamma_{\eta}(E_{nm})=K_\eta(E_{nm})n_\eta(E_{nm})$ and $\tilde{\gamma}_{\eta}(E_{nm})=K_\eta(E_{nm})\left(1+n_\eta(E_{nm})\right)$. Note that, $K_\eta(\epsilon)=0$ for $\epsilon<0$, so only one term survives in the right hand side of Eq.~(\ref{eq:tra_rate}) depending on the choice made for the states $m$ and $n$.

The heat current flowing out of bath C can be expressed as
\begin{multline}
J_{\rm C}=\sum_{l=\pm}\Big[E_l\left(\Gamma_{0l,{\rm C}}p_0-\Gamma_{l0,{\rm C}}p_l\right)\\
+
(E_d-E_l)\left(\Gamma_{ld,{\rm  C}}p_l-\Gamma_{dl,{\rm  C}}p_d\right)\Big].
\end{multline}
\subsection{Local master equation}
\label{app:masloc}
When the $XX$ coupling term ($\Delta_x$) is very small compared to the system-bath coupling strength, local master equation gives a better description of the dynamics. In this case, it is considered that tunneling occurs locally in and out of individual qubits and the energy required for tunneling is independent of $\Delta_x$. The density matrix satisfies a modified Liouville equation. In particular, the diagonal terms of the density matrix satisfy
\begin{equation}
\dot{\rho}_{nn}=-i\left[H_S,\rho\right]_{nn}-\sum_m \Gamma_{nm}\rho_{nn}+\sum_m\Gamma_{m n}\rho_{mm},
\end{equation}
whereas the off-diagonal terms satisfy
\begin{equation}
\dot{\rho}_{mn}=-i\left[H_S,\rho\right]_{mn}-\frac{1}{2}\sum_l\left[\Gamma_{ml}+\Gamma_{nl}\right]\rho_{mn},
\end{equation}
where the transition rates are defined as
\begin{equation}
\Gamma_{mn}=\tilde{\gamma}_\eta(\epsilon_{mn})+\gamma_\eta(\epsilon_{nm}),
\end{equation}
where $\eta$ is the bath associated with the transition. For instance, let's consider the diagonal component $\rho_{11}\equiv p_1$. We have
\begin{multline}
\frac{dp_{1}}{dt}=-i\Delta_x\left(\rho_{21}-\rho_{12}\right)\\
-\left[\Gamma_{10}+\Gamma_{1d}\right]p_{1}+\Gamma_{01}p_{0}+\Gamma_{d1}p_{d}.
\end{multline}
Using the fact that in the steady state $\dot{\rho}_{mn}=0$ for all $m,n$, the component $\rho_{12}$ can be expressed as
\begin{align}
\rho_{12}=\rho_{21}^*=\frac{\Delta_x\left(p_{1}-p_{2}\right)}{\left(\epsilon_1-\epsilon_2\right)-\frac{i}{2}\left(\Gamma_{10}+\Gamma_{1d}+\Gamma_{20}+\Gamma_{2d}\right)},
\end{align}
such that
\begin{multline}
0=\frac{dp_1}{dt}=\frac{\Delta_x^2
\tilde{\Gamma}}{(\epsilon_1-\epsilon_2)^2+\frac{1}{4}\tilde{\Gamma}^2}\left(p_{2}-p_{1}\right)\\
-\left[\Gamma_{10}+\Gamma_{1d}\right] p_{1}+\Gamma_{01}p_{0}+\Gamma_{d1}p_{d},
\end{multline}
where $\tilde {\Gamma}=\Gamma_{10}+\Gamma_{1d}+\Gamma_{20}+\Gamma_{2d}$. Similarly
\begin{multline}
0=\frac{dp_{2}}{dt}=\frac{\Delta_x^2
\tilde{\Gamma}}{(\epsilon_1-\epsilon_2)^2+\frac{1}{4}\tilde{\Gamma}^2}\left(p_{1}-p_{2}\right)\\
-\left[\Gamma_{20}+\Gamma_{2d}\right]p_{2}+\Gamma_{02}p_{0}+\Gamma_{d2}p_{d},
\end{multline}
and
\begin{equation}
0=\frac{dp_0}{dt}=-\left(\Gamma_{01}+\Gamma_{02}\right)p_{0}+\Gamma_{10}p_{1}+\Gamma_{20}p_{2}.
\end{equation}
Along with the condition, $p_0+p_1+p_2+p_d=1$, one can solve the set of master equations.
The heat current flowing out of bath C for the case of local master equation can be expressed as
\begin{equation}
J_{\rm C}=\epsilon_{2}\left(\Gamma_{02}p_0-\Gamma_{20}p_2\right)+(\epsilon_2+\Delta_z)\left(\Gamma_{1d}p_1-\Gamma_{d1}p_d\right).
\end{equation} 
\subsection{Master equations for $\Delta_x=0$}
\label{app:masx0}
For $\Delta_x=0$, both global and local master equations give the same result. We take the local master equations and substitute $\Delta_x=0$. For $p_1$, we obtain
\begin{equation}
0=\frac{dp_1}{dt}=-\left[\Gamma_{10}+\Gamma_{1d}\right] p_{1}+\Gamma_{01}p_{0}+\Gamma_{d1}p_{d}.
\end{equation}
Similarly
\begin{equation}
0=\frac{dp_{2}}{dt}=
-\left[\Gamma_{20}+\Gamma_{2d}\right]p_{2}+\Gamma_{02}p_{0}+\Gamma_{d2}p_{d},
\end{equation}
and
\begin{equation}
0=\frac{dp_0}{dt}=-\left(\Gamma_{01}+\Gamma_{02}\right)p_{0}+\Gamma_{10}p_{1}+\Gamma_{20}p_{2}.
\end{equation}
The heat current flowing out of bath C for the case of local master equation can be expressed as
\begin{equation}
J_{\rm C}=\epsilon_{2}\left(\Gamma_{02}p_0-\Gamma_{20}p_2\right)+(\epsilon_2+\Delta_z)\left(\Gamma_{1d}p_1-\Gamma_{d1}p_d\right).
\end{equation}
\section{Master Equation for non-linear resonator}
\label{app:masres}
We take the following Hamiltonian for the system
\begin{equation}
H_S=\mathbf{1}\otimes\left(\epsilon_1 a_1^\dagger a_1 +\epsilon_2 a_2^\dagger a_2 + \Delta_z a_1^\dagger a_1a_2^\dagger a_2\right),
\end{equation}
where the unit operator $\mathbf{1}$ represents the bath Hilbert space.
The energy eigenvalues can be written in terms of number operators (which commute with the Hamiltonian), ${E}=\epsilon_1 {N}_1+\epsilon_2 {N}_2+\Delta_z {N}_1{N}_2$. The contact Hamiltonian can be similarly written as a tensor product between system and bath degrees of freedom
\begin{multline}
H_C=\sum_n B_n\otimes A_n\\
=\sum_{k,\alpha={\rm L,R}}V_{k\alpha}b_{k\alpha}^\dagger \otimes a_1+ 
\sum_{k}V_{kC}b_{kC}^\dagger \otimes a_1 +h.c.,
\end{multline}
where $A_n$ gives the system degrees of freedom and $B_n$ the bath degrees of freedom such that
\begin{align}
&A_1=a_1~~~~~~~~~~~~~B_1=
b_{kL}^\dagger,\nonumber \\
&A_2=a_1^\dagger~~~~~~~~~~~~~B_2=
b_{kL}\nonumber \\
&A_3=a_1~~~~~~~~~~~~~B_3=
b_{kR}^\dagger,\nonumber \\
&A_4=a_1^\dagger~~~~~~~~~~~~~B_4=
b_{kR}\nonumber \\
&A_5=a_2~~~~~~~~~~~~~B_5=
b_{kC}^\dagger,\nonumber \\
&A_6=a_2^\dagger~~~~~~~~~~~~~B_6=
b_{kC}
\end{align}
Further,
\begin{align}
\frac{da_1}{d\tau}&=-i\epsilon_1 a_1 -i\Delta_z {N}_2a_1,\nonumber \\
\frac{da_1^\dagger}{d\tau}&=i\epsilon_1 a_1^\dagger +i\Delta_z {N}_2a_1^\dagger,\nonumber \\
\frac{da_2}{d\tau}&=-i\epsilon_2 a_2 -i\Delta_z {N}_1a_2,\nonumber \\
\frac{da_2^\dagger}{d\tau}&=i\epsilon_2 a_2^\dagger +i\Delta_z {N}_1a_2^\dagger.
\end{align}
We can write
\begin{align}
&A_1^{\cal H}(\tau)=A_3^{\cal H}(\tau)=e^{-i{\tilde{\omega}}_1\tau}a_1\nonumber \\
&A_2^{\cal H}(\tau)=A_4^{\cal H}(\tau)=e^{i{\tilde{\omega}}_1\tau}a_1^\dagger\nonumber \\
&A_5^{\cal H}(\tau)=e^{-i{\tilde{\omega}}_2\tau}a_2\nonumber \\
&A_6^{\cal H}(\tau)=e^{i{\tilde{\omega}}_2\tau}a_2^\dagger,
\end{align}
where ${\tilde{\omega}}_1=\epsilon_1+\Delta_z {N}_2$ and ${\tilde{\omega}}_2=\epsilon_2+\Delta_z {N}_1$. After some calculations, we obtain\cite{archak}
\begin{widetext}
\begin{multline}
\dot{\rho}(t)=-\int_0^\infty d\tau \bigg[C_{12}(\tau)\Big[A_1,A_2^{\cal H}(-\tau)\rho(t)\Big]+ C_{21}(-\tau)\Big[\rho(t)A_{2}^{\cal H}(-\tau),A_1\Big]+ C_{21}(\tau)\Big[A_2,A_1^{\cal H}(-\tau)\rho(t)\Big]\\
+ C_{12}(-\tau)\Big[\rho(t)A_{1}^{\cal H}(-\tau),A_2\Big]   
+C_{34}(\tau)\Big[A_3,A_4^{\cal H}(-\tau)\rho(t)\Big]+ C_{43}(-\tau)\Big[\rho(t)A_{4}^{\cal H}(-\tau),A_3\Big]+ C_{43}(\tau)\Big[A_4,A_3^{\cal H}(-\tau)\rho(t)\Big]\\
+ C_{34}(-\tau)\Big[\rho(t)A_{3}^{\cal H}(-\tau),A_4\Big]
+C_{56}(\tau)\Big[A_5,A_6^{\cal H}(-\tau)\rho(t)\Big]+ C_{65}(-\tau)\Big[\rho(t)A_{6}^{\cal H}(-\tau),A_5\Big]+ C_{65}(\tau)\Big[A_6,A_5^{\cal H}(-\tau)\rho(t)\Big]\\
+ C_{56}(-\tau)\Big[\rho(t)A_{5}^{\cal H}(-\tau),A_6\Big]\bigg],
\label{eq:eom_lin}
\end{multline}
\end{widetext}
where $C_{\alpha\beta}(\tau)={\rm Tr}_{\rm B}\left[B_{\alpha}(\tau)B_\beta\right]$. We took the only combinations for $\alpha$ and $\beta$ that give $C_{\alpha\beta}\neq 0$. We defined
\begin{equation}
\Gamma_{\alpha\beta}(\omega)=\int_0^\infty C_{\alpha\beta}(\tau)e^{i\hat{\tilde{\omega}}\tau},
\label{eq:gam_albe}
\end{equation}
and
\begin{equation}
\Gamma_{\alpha\beta}^*({\tilde{\omega}})=\int_0^\infty d\tau C_{\alpha\beta}(-\tau)e^{-i{\tilde{\omega}}\tau}.
\label{eq:gamst_albe}
\end{equation}
Since the contact Hamiltonian is Hermitian, we have $C_{\alpha\beta}(-\tau)=C_{\alpha\beta}^\dagger(\tau)$. Further, we have the following relation
\begin{equation}
\Gamma_{\alpha\beta}(\omega)+\Gamma_{\alpha\beta}^*(\omega)=\int_{-\infty}^\infty d\tau C_{\alpha\beta}(\tau)e^{i\omega \tau}.
\end{equation}
The real part of $\Gamma$'s give the dissipative contribution whereas the imaginary part gives the Lamb shift. Substituting Eqs.~(\ref{eq:gam_albe}) and (\ref{eq:gamst_albe}) in Eq.~(\ref{eq:eom_lin}) we obtain
\begin{widetext}
\begin{multline}
\dot{\rho}=-\bigg[\Big[a_1,\Gamma_{12}(-{\tilde{\omega}}_1)a_1^\dagger\rho(t)\Big]+\Big[\rho(t)\Gamma_{21}^*({\tilde{\omega}}_1)a_1^\dagger,a_1\Big]+\Big[a_1^\dagger,\Gamma_{21}({\tilde{\omega}}_1)a_1\rho(t)\Big]+\Big[\rho(t)\Gamma_{12}^*(-{\tilde{\omega}}_1)a_1,a_1^\dagger\Big]\\
+\Big[a_1,\Gamma_{34}(-{\tilde{\omega}}_1)a_1^\dagger\rho(t)\Big]+\Big[\rho(t)\Gamma_{43}^*({\tilde{\omega}}_1)a_1^\dagger,a_1\Big]+\Big[a_1^\dagger,\Gamma_{43}({\tilde{\omega}}_1)a_1\rho(t)\Big]+\Big[\rho(t)\Gamma_{34}^*(-{\tilde{\omega}}_1)a_1,a_1^\dagger\Big]\\+\Big[a_2,\Gamma_{56}(-{\tilde{\omega}}_1)a_2^\dagger\rho(t)\Big]+\Big[\rho(t)\Gamma_{65}^*({\tilde{\omega}}_2)a_2^\dagger,a_2\Big]+\Big[a_2^\dagger,\Gamma_{65}({\tilde{\omega}}_1)a_2\rho(t)\Big]+\Big[\rho(t)\Gamma_{56}^*(-{\tilde{\omega}}_2)a_2,a_2^\dagger\Big]\bigg].
\end{multline}
Keeping only the dissipative contributions, we can write the above expression as
\begin{multline}
\dot{\rho}=-\sum_{\alpha={\rm L,R}}\bigg[\Big[a_1,{\cal F}_\alpha({\tilde{\omega}}_1)a_1^\dagger\rho(t)\Big]+\Big[\rho(t){\cal G}_\alpha({\tilde{\omega}}_1)a_1^\dagger,a_1\Big]+\Big[a_1^\dagger,{\cal G}_\alpha({\tilde{\omega}}_1)a_1\rho(t)\Big]+\Big[\rho(t){\cal F}_\alpha({\tilde{\omega}}_1)a_1,a_1^\dagger\Big]\bigg]\\
-\bigg[\Big[a_2,{\cal F}_{\rm C}({\tilde{\omega}}_2)a_2^\dagger\rho(t)\Big]+\Big[\rho(t){\cal G}_{\rm C}({\tilde{\omega}}_2)a_2^\dagger,a_2\Big]+\Big[a_2^\dagger,{\cal G}_{\rm C}({\tilde{\omega}}_2)a_2\rho(t)\Big]+\Big[\rho(t){\cal F}_{\rm C}({\tilde{\omega}}_2)a_2,a_2^\dagger\Big]\bigg],
\end{multline}
where
\begin{align}
&{\cal F}_{\rm L}({\tilde{\omega}})={\rm Re}
[\Gamma_{12}(-{\tilde{\omega}})]={\rm Re}
[\Gamma_{12}^*(-{\tilde{\omega}})]=\frac{1}{2} K_{\rm L}({\tilde{\omega}})
n_{\rm L}({\tilde{\omega}})\nonumber \\
&{\cal G}_{\rm L}({\tilde{\omega}})={\rm Re}
[\Gamma_{21}^*({\tilde{\omega}})]={\rm Re}
[\Gamma_{21}({\tilde{\omega}})]=\frac{1}{2} K_{\rm L}({\tilde{\omega}})
(1+n_{\rm L}
({\tilde{\omega}})),\nonumber \\
&{\cal F}_{\rm R}({\tilde{\omega}})={\rm Re}
[\Gamma_{34}(-{\tilde{\omega}})]={\rm Re}
[\Gamma_{34}^*(-{\tilde{\omega}})]=\frac{1}{2} K_{\rm R}({\tilde{\omega}})
n_{\rm R}({\tilde{\omega}})\nonumber \\
&{\cal G}_{\rm R}({\tilde{\omega}})={\rm Re}
[\Gamma_{43}^*({\tilde{\omega}})]={\rm Re}
[\Gamma_{43}({\tilde{\omega}})]=\frac{1}{2} K_{\rm R}({\tilde{\omega}})
(1+n_{\rm R}
({\tilde{\omega}})),  
\nonumber \\
&{\cal F}_{\rm C}({\tilde{\omega}})={\rm Re}
[\Gamma_{56}(-{\tilde{\omega}})]={\rm Re}
[\Gamma_{56}^*(-{\tilde{\omega}})]=\frac{1}{2} K_{\rm C}({\tilde{\omega}})
n_{\rm C}({\tilde{\omega}})\nonumber \\
&{\cal G}_{\rm C}({\tilde{\omega}})={\rm Re}
[\Gamma_{65}^*({\tilde{\omega}})]={\rm Re}
[\Gamma_{65}({\tilde{\omega}})]=\frac{1}{2} K_{\rm C}({\tilde{\omega}})
(1+n_{\rm C}
({\tilde{\omega}})),  
\end{align}
where $K_\eta(\omega)$ and $n_\eta(\omega)$ are respectively the spectral density and the distribution function for the bath $\eta$. After some calculations we obtain
\begin{multline}
\dot{\rho}_{n_1n_2}=-2\sum_{\alpha={\rm L,R}}\left[(n_1+1){\cal F}_\alpha(\tilde{\omega}_{1,n_2})\rho_{n_1,n_2}-n_1{\cal F}_\alpha(\tilde{\omega}_{1,n_2})\rho_{n_1-1,n_2}+n_1 G_\alpha(\tilde{\omega}_{1,n_2})\rho_{n_1,n_2}-(n_1+1)G_\alpha(\tilde{\omega}_{1,n_2})\rho_{n_1+1,n2}\right]\\
-2\left[(n_2+1){\cal F}_{\rm C}(\tilde{\omega}_{2,n_1})\rho_{n_1,n_2}-n_2{\cal F}_{\rm C}(\tilde{\omega}_{2,n_1})\rho_{n_1,n_2-1}+n_2 G_{\rm C}(\tilde{\omega}_{2,n_1})\rho_{n_1,n_2}-(n_2+1)G_{\rm C}(\tilde{\omega}_{2,n_1})\rho_{n_1,n2+1}\right],
\end{multline}
\end{widetext}
where $\tilde{\omega}_{1,n_2}=\epsilon_1+\Delta_z n_2$ and $\tilde{\omega}_{2,n_1}=\epsilon_2+\Delta_z n_1$. Defining
\begin{align}
&C_1({n_1},{n_2})=2\sum_\alpha(n_1+1){\cal }F_\alpha(\tilde
{\omega}_{1,n_2}),
\nonumber\\
&C_2({n_1},{n_2})=2(n_2+1){\cal F}_{\rm C}(\tilde
{\omega}_{2,n_1})\nonumber \\
&D_1({n_1},{n_2})=2\sum_\alpha n_1{\cal G}_\alpha(\tilde
{\omega}_{1,n_2}),\nonumber \\
&D2({n_1},{n_2})=2n_2{\cal G}_{\rm C}(\tilde
{\omega}_{2,n_1})
\end{align}
we obtain
\begin{multline}
\dot{\rho}_{n_1,n_2}=-\bigg[\Big(C_1({n_1},{n_2})+D_1({n_1},{n_2})+C_2({n_1},{n_2})\\
+D_2({n_1},{n_2})\Big)\rho_{n_1,n_2}-C_1({n_1-1},{n_2})\rho_{n_1-1,n_2}-D_1({n_1+1},{n_2})\\
\rho_{n_1+1,n_2}-C_2(n1,{n_2-1})\rho_{n_1,n_2-1}-D_2(n1,{n_2+1})\rho_{n_1,n_2+1}\bigg].
\end{multline}
Hence we derived the master equation for coupled resonators.
Similarly, the heat current flowing into the bath C can be written as
\begin{equation}
J_{\rm C}=-{\rm Tr}\left[H_S{\cal L}_{\rm C}\rho\right],
\end{equation}
where ${\cal L}_{\rm C}$ is the part of Lindbladian associated with the bath C given by
\begin{multline}
[{\cal L}_{\rm C} \rho]_{n_1,n_2}=-\bigg[\Big(C_2({n_1},{n_2})
+D_2({n_1},{n_2})\Big)\rho_{n_1,n_2}\\
-C_2(n_1,{n_2-1})\rho_{n_1,n_2-1}-D_2(n_1,{n_2+1})\rho_{n_1,n_2+1}\bigg].
\end{multline}
\section{Two coupled non-linear resonators: Dyson equation in the mean field approximation}
\label{app:greenres}
We define the time-ordered Green's function for the coupled resonator system as
\begin{equation}
G_{i;j}(t,t')=-i\left\langle{\cal T} a_i(t)a_{j}^\dagger(t')\right\rangle.
\end{equation}
Using the equation of motion, we obtain
\begin{equation}
i\partial_t G_{i;j}(t,t')=-i\left\langle{\cal T}\left[a_i(t),H(t)\right]a_j^\dagger(t')\right\rangle +\delta_{ij}\delta(t-t'),
\label{eq:dyson01}
\end{equation}
where $i,j=1,2$. For the commutator, we have
\begin{equation}
\left[a_i,H\right]=\epsilon_ia_i+\sum_k V_{ki}b_{ki}+\Delta_z n_l a_i,
\label{eq:commutat}
\end{equation}
where $n_i=a_i^\dagger a_i$ and $l\neq i$. Substituting (\ref{eq:commutat}) in (\ref{eq:dyson01}), we obtain
\begin{multline}
i\partial_t G_{i;j}(t,t')=\delta_{ij}\delta(t-t')+\epsilon_i G_{i;j}(t,t')\\
+\Delta_z G_{n_l i;j}(t,t') +\sum_{k,\eta} V_{k\eta} G_{k\eta;j}(t,t')
,
\label{eq:dyson02}
\end{multline}
where $\eta={\rm L,R}$ for $i=1$ and $\eta={\rm C}$ for $i=2$. $G_{n_i i;j}(t,t')=-i\left\langle{\cal T}n_i(t)a_i(t)a_j^\dagger(t')\right\rangle$ and $G_{k\eta;j}(t,t')=-i\left\langle{\cal T} b_{k\gamma}(t)a_j^\dagger(t')\right\rangle$. Using equation of motion it is straightforward to show that
\begin{equation}
G_{k\eta;j}(t,t')=V_{k\eta}\int dt_1 g_{k\eta;k\eta}(t,t_1)G_{i;j}(t_1,t'),
\label{eq:dyson03}
\end{equation}
where $g_{k\eta;k\eta}$ is the time-ordered Green's function for the free bath. Substituting Eq.~(\ref{eq:dyson03}) in Eq.~ 
(\ref{eq:dyson02}), we obtain
\begin{multline}
i\partial_t G_{i;j}(t,t')=\delta_{ij}\delta(t-t')+\epsilon_i G_{i;j}(t,t')\\
+\Delta_z G_{n_l i;j}(t,t')+\sum_\eta\int dt_1 \Sigma_{\eta}^{(1)}(t,t_1)G_{i;j}(t_1,t'),
\end{multline}
where $\Sigma_\eta(t_1,t_2)=\sum_{k\eta}|V_{k\eta}|^2g_{k\eta}(t_1,t_2)$ is the embedded self-energy due to system-bath coupling. Under {\em the mean field approximation}
\begin{multline}
i\partial_t G_{i;j}(t,t')=\delta_{ij}\delta(t-t')+\epsilon_i G_{i;j}(t,t')+\Delta_z \left\langle n_l\right\rangle G_{ i;j}(t,t')\\
+\sum_\eta\int dt_1 \Sigma_{\eta}(t,t_1)G_{i;j}(t_1,t').
\end{multline}
The Dyson equation for the retarded Green's function can be obtained simply by replacing time-ordered Green's functions and self energies with the retarded counterpart.

\bibliography{paper}

\end{document}